\documentclass{article}

\usepackage[preprint]{neurips_2026}
\usepackage[utf8]{inputenc} 
\usepackage[T1]{fontenc}    
\usepackage{hyperref}       
\usepackage{url}            
\usepackage{booktabs}       
\usepackage{amsfonts}       
\usepackage{nicefrac}       
\usepackage{microtype}      
\usepackage{xcolor}         
\usepackage{wrapfig}

\usepackage{microtype}
\usepackage{cleveref}
\usepackage{xspace}
\usepackage{graphicx}
\usepackage{subcaption}
\usepackage{wrapfig}
\usepackage{makecell}
\usepackage{lipsum}  
\usepackage{subcaption}
\usepackage{verbatim}
\usepackage{enumitem}

\usepackage{array}
\newcommand{\PreserveBackslash}[1]{\let\temp=\\#1\let\\=\temp}
\newcolumntype{C}[1]{>{\PreserveBackslash\centering}p{#1}}
\newcolumntype{R}[1]{>{\PreserveBackslash\raggedleft}p{#1}}
\newcolumntype{L}[1]{>{\PreserveBackslash\raggedright}p{#1}}

\definecolor{lue}{RGB}{135, 206, 250}
\definecolor{green}{RGB}{185, 226, 207}

\newcommand{\eg}{e.g.,\xspace}
\newcommand{\ie}{i.e.\xspace}
\newcommand{\metricname}{\emph{offloading score}\xspace}
\newcommand{\Metricname}{\emph{Offloading Score}\xspace}
\newcommand{\outputuse}{\emph{output-use}\xspace}

\newcommand{\cogprocess}{\emph{process}\xspace}
\newcommand{\cogprocesses}{\emph{processes}\xspace}

\newcommand{\shortcond}{\emph{short}\xspace}

\newcommand{\longcond}{\emph{long}\xspace}

\definecolor{lightblue}{rgb}{.8,.8,1}
\definecolor{lightred}{rgb}{1, 0.7, 0.7}

\definecolor{aiassistbg}{HTML}{E6F4F1}
\definecolor{aiassisttext}{HTML}{1F6F67}

\def\shownotes{1}  
\ifnum\shownotes=1
\newcommand{\authnote}[2]{[#1: #2]}
\else
\newcommand{\authnote}[2]{}
\fi
\newcommand{\subtitle}[1]{%
  \posttitle{%
    \par\end{center}
    \begin{center}\large#1\end{center}
    \vskip0.5em}%
}

\def\arxiv{1}
\ifnum\arxiv=1

\else

\fi

\begin{document}

\title{Offloading Score: Measuring AI Reliance Through Counterfactual Workflows}

\author{
Vishakh Padmakumar \\
Stanford University \\
\And
Lujain Ibrahim \\
University of Oxford \\
\And
Zora Zhiruo Wang \\
Carnegie Mellon University \\
\And
Jennifer Wang \\
Stanford University \\
\And
Q. Vera Liao \\
University of Michigan \\
\And
Diyi Yang \\
Stanford University \\
}
\maketitle

\begin{abstract}
AI tools are increasingly integrated into real-world workflows. However, existing measures of reliance on these tools focus on AI output adoption or on self-reported indicators, rather than how task effort is distributed between users and tools. 
Here, we introduce \metricname, a measure of reliance that quantifies the fraction of cognitive effort offloaded to an AI tool. \Metricname is simulation-based---we construct a counterfactual workflow by estimating how the user would have completed the task without the tool, and then computing the fraction of steps saved by using the tool. 
We validate \metricname through intrinsic evaluations of metric validity, and a controlled user study ($n=40$) with developers performing programming tasks using AI tools. 
We vary time pressure to test whether reliance measures capture the known increase in reliance under time pressure.
We show that \metricname detects significantly higher reliance in time-constrained settings ($+43\%$, $p=0.018$), while usage-based and self-reported baseline measures of reliance do not distinguish the conditions. 
We complement this with descriptive insights showing that higher reliance manifests as greater delegation of subtasks to the tool and more direct reuse of AI outputs.
Finally, we demonstrate an approach of using \metricname in combination with target outcomes of a task (e.g., code understanding) to identify when reliance may be (in)appropriate. Our framework offers two contributions: an instrument users can apply to measure and reflect on their own reliance, and a quantitative signal that agent designers can utilize to mitigate overreliance.
\end{abstract}



\section{Introduction}
\label{sec:intro}

AI tools are increasingly integrated into everyday cognitive tasks, from writing and coding to analysis and decision-making. 
While these tools can improve productivity \citep{brynjolfsson2025generative} and are thus being widely adopted \citep{sleegers2025adoption}, they also raise concerns about risks of overreliance, including deskilling and reduced independent problem-solving ability \citep{shukla2025skilling, zhi2026investigating, shen2026ai, ibrahim2025measuring}.
These concerns extend beyond the quality of output produced, to cognitive impacts resulting from offloading of cognitive work to AI.

Existing frameworks typically measure reliance either using users’ self-assessments or AI output adoption (\Cref{sec:background}). Output-based methods, developed in prior work focusing on classification settings, treat reliance as a binary measurement (accept/reject), categorizing users as overreliant when they accept incorrect outputs and underreliant when they reject correct ones \citep{vasconcelos2023explanations, liu2026behavioral}. These measures break down 
in multi-turn human-AI workflows with contemporary AI tools, 
where reliance may manifest in different ways users provide inputs (i.e., prompting) and interact with tool outputs.
Self-reported measures are more fine-grained in capturing users’ perceived reliance, but are subjective, noisy, and expensive to collect \citep{kohn2021measurement}. 

\begin{figure}
    \centering
    \includegraphics[width=\linewidth]{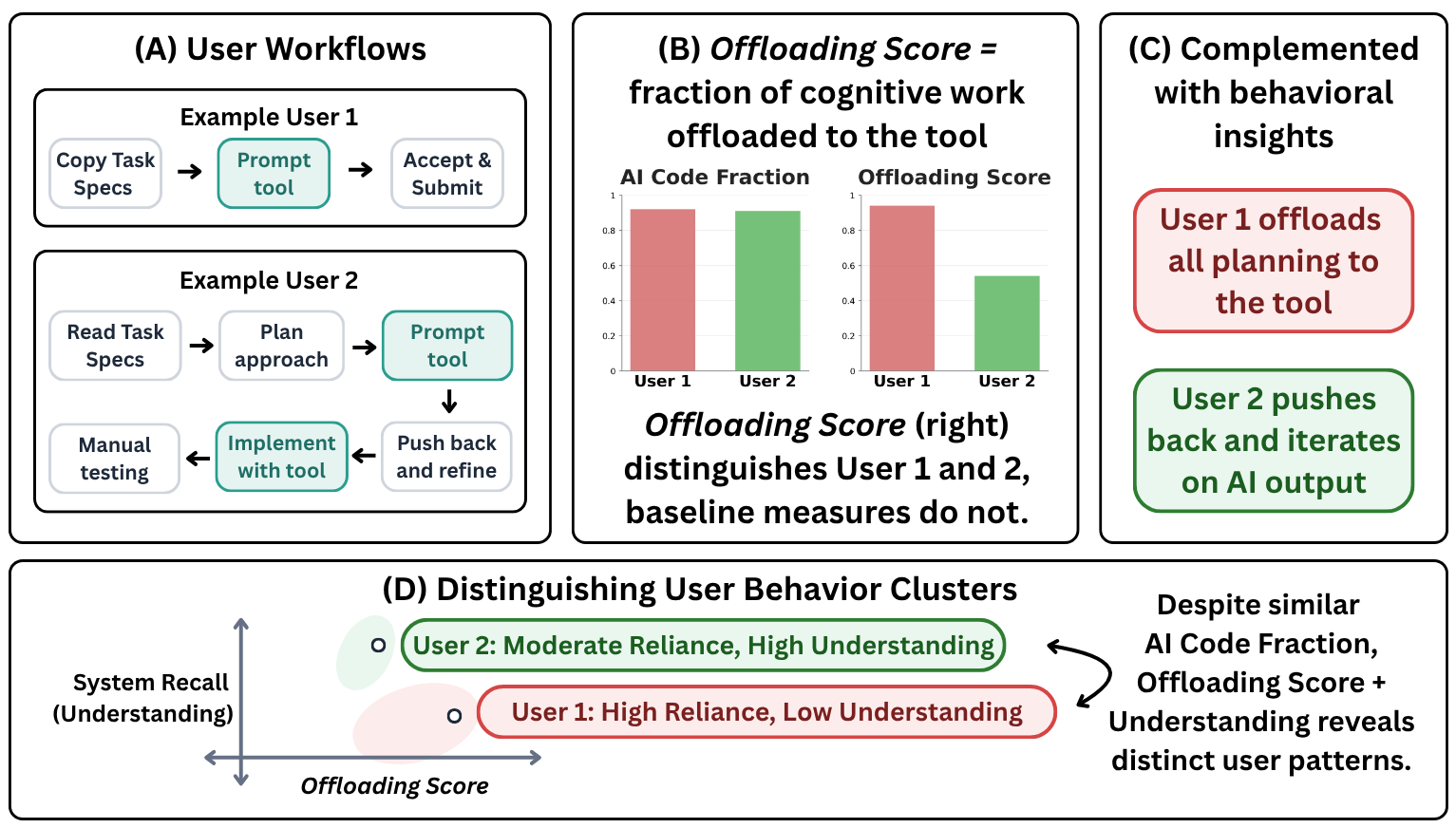}
    \caption{
    We propose \metricname, a scalar measure of cognitive effort offloaded to an AI tool, complemented by descriptive dimensions of tool use. \Metricname is computed by identifying \colorbox{aiassistbg}{\textcolor{aiassisttext}{AI-assisted}} workflow steps, estimating human-only counterfactual alternatives, and computing the fraction of workflow steps saved by the tool. \Metricname distinguishes user reliance patterns more effectively than baseline measures (\eg AI Code Fraction) (\Cref{sec:validation_user_study}) and can be paired with outcome measures such as system recall to interpret reliance patterns (\Cref{sec:normative_judgment}).
    }
    \label{fig:fig1}
\end{figure}

In this work, we instead characterize reliance by how cognitive work is distributed between the user and the AI tool, \ie how many of the planning, execution, and verification steps that a user would otherwise perform are offloaded to the AI. Intuitively, a user who asks the tool to solve an entire coding task is more reliant than one who decomposes the task, queries the model for subcomponents, and verifies each step, even if both produce similar final outputs (\Cref{fig:fig1} (A)). Our goal is to measure how AI use changes people's cognitive processes when completing tasks, and provide a starting point for identifying inappropriate reliance patterns that may lead to negative long-term consequences. 

\paragraph{RQ1: Can we design a measure of reliance that reflects what and how much cognitive work is offloaded to AI?}
We propose a multi-dimensional framework for measuring reliance (\Cref{sec:formulation}, \Cref{fig:formulation_dimensions}): (1) A primary scalar metric, \metricname, that estimates the fraction of cognitive effort offloaded to the tool by replacing each AI-assisted workflow step with a human-only counterfactual sequence and measuring the fraction of steps saved (\Cref{fig:formulation_offloading_score}); and (2) complementary descriptive dimensions that categorize (a) the kinds of tasks offloaded to the tool, inspired by the Flower model of cognitive processes \citep{flower1981cognitive}, and (b) how users engage with model outputs, using a four-point scale inspired by Bloom's taxonomy \citep{bloom1956taxonomy}.
We first establish the validity of \metricname by evaluating whether the constructed human counterfactual workflows provide reasonable alternatives for completing the same task, as well as assessing the stability and sensitivity of the metric to controlled perturbations (\Cref{sec:pre_validation_metrics_validity}). 
We then validate \metricname in a controlled user study with $N = 40$  experienced freelance developers completing programming tasks either under \shortcond ($1$ hour) or \longcond ($4$ hours) time limits. To reflect the association between time pressure and increased tool reliance \citep{zakay1993impact, rice2009automation, swaroop2024accuracy, haduong2024performance}, a valid reliance measure should assign higher scores in the \shortcond condition. We find that \metricname assigns significantly higher reliance in the \shortcond condition (on average $0.451$ vs. $0.308$, $p=0.018$), while baseline measures, including output-based metrics such as fraction of AI-code retained ($0.152$ vs $0.052$, $p=0.072$) and self-reported cognitive load ($3.63$ vs. $3.56$, $p=0.881$), do not significantly distinguish the conditions (\Cref{sec:user_study_results}).
We also identify via the descriptive dimensions that the higher reliance is associated with a shift in usage patterns, leading to more direct reuse of AI outputs and less iterative back-and-forth with the tool.

\paragraph{RQ2: Can we use \metricname to identify conditions for (in)appropriate reliance?}
The judgment of whether a reliance level should be considered appropriate depends on the user’s normative goal, i.e., what outcome reliance is meant to support.
We demonstrate an approach to use \metricname in combination with measures of target task outcome, using a desirable level of code understanding as an example, to identify the conditions for appropriate reliance. We observe a general negative correlation between \metricname and task understanding, which allows setting a threshold of \metricname for overreliance---higher-than-threshold reliance leading to unacceptably poor task understanding. However, we also observe an outlier cluster of users who have moderate-to-high \metricname and high task understanding. Qualitative data indicate that they exhibit a distinct pattern of using the AI tool for \textit{learning} about unfamiliar coding tasks, suggesting higher reliance can be considered \emph{appropriate} in this particular context of AI tool use.

Concretely, we sum up our contributions as:
    (1) \textbf{A measure of reliance grounded in user behavior.} We introduce \metricname, a scalar measure that estimates the fraction of workflow steps saved through AI assistance by replacing AI-assisted steps with human-only counterfactual alternatives. \Metricname is computed directly from interaction traces (\eg screenshots and keystrokes), making it applicable across different tools and interfaces.
    (2) \textbf{A multi-dimensional characterization of AI usage.} We complement \metricname with descriptive categorization of the tasks offloaded to AI and how users engage with AI outputs.  
    (3) \textbf{Empirical validation through a user study.} \Metricname captures variation in reliance under time pressure better than baseline measures, while providing descriptive behavioral insight.
    By jointly analyzing reliance and code understanding, we show that users exhibit distinct patterns, including overreliance and appropriate reliance.\footnote{We release the \href{https://github.com/vishakhpk/offloading-score}{code} associated with this project as well as a \href{https://vishakhpk.github.io/measuring-reliance/}{website} with examples and consolidated findings.}
\section{Background}
\label{sec:background}
\textbf{Existing measures of reliance}
Prior work has measured reliance using two main approaches. The first is \textit{usage-based}, usually in the form of a binary value measuring whether or not users adopt AI-provided outputs \citep{vasconcelos2023explanations, buccinca2021trust, zhou2025rel, bansal2021does}. These measures are often combined with evaluations of outcome correctness to assign normative labels such as overreliance (adopt + incorrect) and underreliance (reject + correct) to an interaction. Other related usage-based metrics, such as switch fraction \citep{yin2019understanding} and weight of advice \citep{logg2019algorithm}, measure whether users update their decisions with AI suggestions. 
The second approach involves self-reported measures of perceived reliance and its cognitive effects such as automation complacency scales \citep{merritt2019automation} and cognitive load assessments (e.g., NASA TLX \citep{hart1988development}). 
\citet{zhi2026investigating} infer reliance from performance differences across AI access conditions, treating it as a latent factor. 
In this work, we propose a measure that bridges these different approaches by going beyond \emph{how much} a tool is used to capture \emph{how} it is used in a workflow, providing a process-oriented view of reliance grounded in user behavior.

\noindent
\textbf{Impacts of (over)reliance}
The public-facing deployment of LLM-based systems at scale has motivated work on the impacts of (over)reliance, characterized by reduced scrutiny of AI outputs, weaker human oversight, and difficulty maintaining appropriate control over AI-supported decisions \citep{passi2022overreliance, buccinca2021trust, yizhou2026understanding}. Recent work extends this concern to longer-term effects, including cognitive surrender, disempowerment, misplaced responsibility, and reduced opportunities for skill development \citep{shaw2026thinking, sharma2026s, shukla2025skilling}. In educational settings, reliance patterns are shaped by user factors such as expertise and need for cognition
\citep{pitts2025students, shen2026ai}. 
These concerns are salient in AI-assisted coding, one of the tasks and professions most exposed to contemporary models \citep{massenkoffmccrory2026labor}. Coding assistants reduce effort and automate unfamiliar tasks but also raise concerns about whether users can understand and maintain the code they produce \citep{chen2025code}. AI tools may slow experienced developers in realistic open-source tasks, and AI-generated code can increase maintenance burden and technical debt when developers must later understand, review, or repair generated artifacts \citep{becker2025measuring, xu2026ai}. These findings motivate our focus on capturing how users delegate cognitive effort to AI systems and how well users understand the resulting code artifacts, beyond just the task outcome.
\vspace{-2mm}
\section{Process-Oriented Measurement of Reliance}
\label{sec:formulation}
\vspace{-2mm}
\subsection{Problem Setting}
\label{sec:problem_setting}
\vspace{-2mm}
We consider the problem of characterizing the reliance of a user $U$ on a tool $T$ when performing computer-use activities. Following \citet{wang2025ai}, we define a \emph{workflow} as a sequence of steps taken to achieve a predefined goal, where each step consists of one or more actions that accomplish a distinguishable sub-goal. Formally, let $W = \{w_1, \dots, w_n\}$ denote a workflow of $n$ steps undertaken by $U$ to achieve a goal. 
Each step $w_i$ represents a coherent unit of progress toward the goal, such as writing code to fix a feature or generating documentation files.
Each step may be completed independently by $U$, left entirely to $T$, or involve interaction between the two for that particular sub-goal. 
Our goal is to characterize how $T$ is used at each step, as well as how the user responds to the outputs of $T$, capturing the reliance of $U$ on $T$ by modeling both the allocation of cognitive effort and the interaction dynamics between them over the course of the workflow $W$.

\begin{figure}
    \centering
    \includegraphics[width=0.85\linewidth]{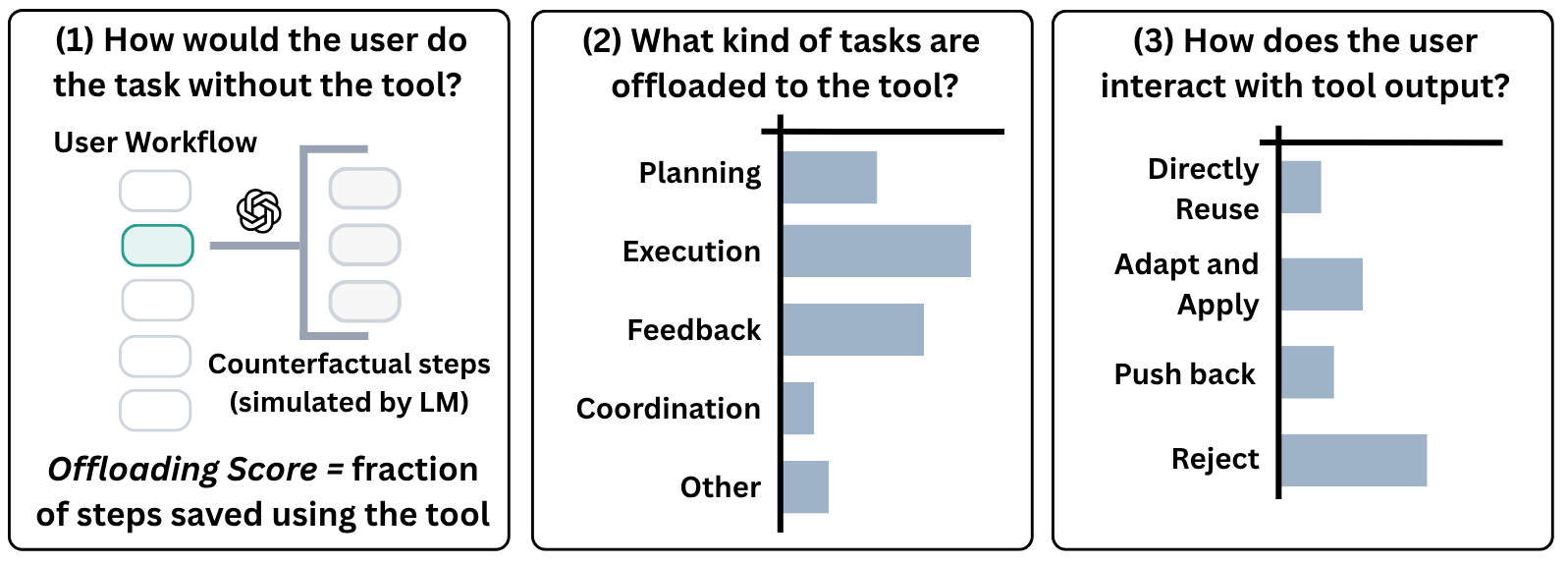}
    \caption{(1) We compute \metricname by expanding AI-assisted steps into simulated counterfactual human-only steps and measuring the fraction saved. \Cref{fig:formulation_offloading_score} details how \metricname is computed. We also characterize reliance by the (2) types of \cogprocesses offloaded and (3) how users interact with tool outputs.}
    \label{fig:formulation_dimensions}
\end{figure}

\subsection{Proposed Measure}
\label{sec:proposed_measure} 
\begin{figure}
    \centering
    \includegraphics[width=0.9\linewidth]{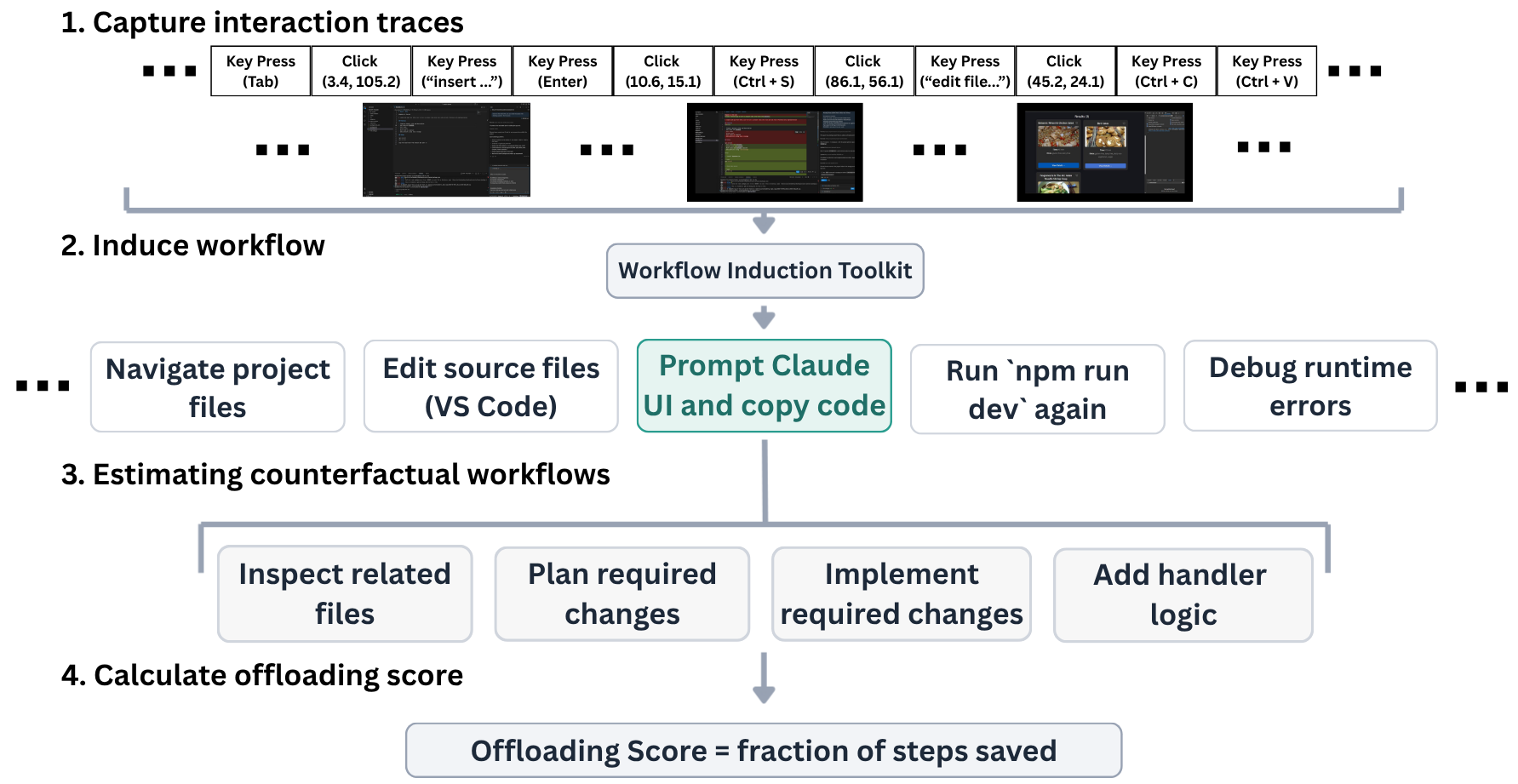}
    \caption{Calculating \metricname. From raw interaction traces (e.g., keystrokes, clicks, screenshots), we induce the observed human–AI workflow using the workflow induction toolkit \citep{wang2025ai}. \colorbox{aiassistbg}{\textcolor{aiassisttext}{AI-assisted steps}} are expanded into equivalent human-only counterfactual sequences, representing how the same sub-goal would be achieved without the tool. We compute the \metricname as the fraction of workflow steps saved through tool use.}
    \label{fig:formulation_offloading_score}
\end{figure}
We first define a scalar measure of reliance based on how cognitive effort in the workflow is distributed between the $U$ and $T$ and introduce complementary dimensions that capture which cognitive processes $U$ offloads to $T$ and how they engage with the tool output (\Cref{fig:formulation_dimensions}). In \Cref{sec:validation_user_study}, we show how these \cogprocess and \outputuse labels provide interpretable, descriptive insights into user behavior.

\paragraph{How would the user have done the task without AI?}
We adopt a counterfactual formulation to capture not just how much the tool is used, but how it is used within the workflow. Since cognitive effort is not directly observable from the final artifacts or tool usage alone, we estimate how each AI-assisted sub-goal would have been completed without the tool.
For each observed workflow $W = \{w_1, \dots, w_n\}$, we identify the subset of steps that were completed with assistance from $T$. For each such step $w_i$, we estimate a corresponding human-only counterfactual sequence $w'_i = \{w'_{i,1}, \dots, w'_{i,k_i}\}$ with $k_i \geq 1$, representing how an average user would have completed the same sub-goal without access to the tool.

We then obtain a counterfactual workflow $W' = \{w'_1, \dots, w'_m\}$ by replacing each AI-assisted step $w_i$ with its corresponding human-only sequence $w'_i$. Typically, $m \geq n$, reflecting the additional work required without the tool. See \Cref{fig:formulation_offloading_score} for an illustrative example. This short-horizon simulation is inspired by past work which simulates interaction trajectories to estimate the contribution of individual steps in a workflow \citep{wu2025collabllm}. Note $W'$ is defined with respect to an average user rather than the specific user $U$, allowing this framework to be reused across different users without requiring a personalized estimate for each individual.\footnote{We note a natural extension for personalized measurement would construct user-specific counterfactuals conditioned on inferred capabilities or prior behavior \citep{shaikh2025creating}.} 

We quantify the cognitive effort offloaded to $T$ by the scalar fraction $\frac{m - n}{m}$, which measures the proportion of counterfactual workflow steps saved through the use of the tool.\footnote{We note that our work assumes workflow induction that produces a consistent granularity across steps as part of the analysis pipeline. We acknowledge some limitations of this process in \Cref{sec:limitations}.} In general, $m \geq n$, so the range of this value is between $0$ and $1$. This fraction, referred to as \metricname, serves as the primary scalar measure of reliance in empirical evaluation. We evaluate how this measure correlates with reference labels of reliance in \Cref{sec:user_study_results} and also how we can use this to distinguish different patterns of user behavior in \Cref{sec:normative_judgment}. 

To complement \metricname, we introduce the following dimensions that provide a structured description of how this reliance manifests. 

\paragraph{What kind of assistance is the user $U$ asking of the tool $T$?}
For each user request, we examine the interaction in context and classify it into a set of cognitive process categories inspired by \citet{flower1981cognitive}. These include \textbf{high-level planning}, \textbf{low-level execution}, \textbf{feedback seeking}, \textbf{coordination}, and an ``other'' category. Each interaction is associated with a workflow step $w_i$, allowing us to characterize the distribution of process types across the workflow. We compute the fraction of workflow steps that involve each type of cognitive process, capturing what kinds of work are being offloaded to the tool. These are referred to as \cogprocess labels.

\paragraph{How is the user engaging with the output of the model?}
In addition to the requests made to the tool, we characterize how users engage with model outputs. Drawing on \citet{bloom1956taxonomy}, we define a set of interaction levels that reflect increasing degrees of user involvement: \textbf{directly reuse} (the user directly incorporates model-generated content with minimal modification), \textbf{adapt and apply} (the user applies ideas or structures from the model output but substantially adapts them), \textbf{question, debug, or push back} (the user evaluates or challenges the model output), and \textbf{reject} (the user does not use the model output and completes the step independently).
We compute the fraction of workflow steps $w_i$ associated with output interactions of each category, capturing how users allocate effort when engaging with tool-generated content. These are referred to as \outputuse labels.

\subsection{Intrinsic Evaluation of \Metricname}
\label{sec:pre_validation_metrics_validity}

We first evaluate \metricname through intrinsic tests of metric validity.

\textbf{Content Validity.} Content validity assesses whether a metric captures the construct it is intended to measure. For constructs that are difficult to quantify directly, we follow prior work on automatic metric design \citep{ryan2025autometrics} by making the metric computation transparent and supporting its components with evidence. \Metricname affords transparency by design as it is computed as a simple fraction of observed and counterfactual workflow steps, and users can review the mapping from each AI-assisted step to its human-only counterfactual. 
We validate the plausibility of these counterfactuals in two ways: (1) through human judgments, with over $85\%$ of sampled counterfactual steps rated plausible ($\geq 3$ on a $5$-point Likert scale)  (\Cref{sec:app_human_validation_counterfactual}); and (2) using a dataset of human and agent workflows from \citet{wang2025ai}, showing that generated counterfactual workflows align more closely with human workflows from the same task than with workflows from other related tasks in the same domain ($p \approx 10^{-6}$ with a Wilcoxon test; $p < 10^{-4}$ with a permutation test) (\Cref{sec:app_validate_human_counterfactual}). 
We also validate the descriptive annotations against human judgments, finding that \texttt{gpt-5.2} achieves $80\%$ agreement for cognitive process labels and $81\%$ agreement for output-use labels (\Cref{sec:app_annotation_llm_as_judge}).

\textbf{Construct Validity.} Construct validity asks whether a metric behaves consistently with the construct it is intended to measure. Following \citet{campbell1959convergent}, we evaluate this through sensitivity and stability. Sensitivity tests whether a metric assigns systematically lower (or higher) scores when examples are perturbed in a manner meant to degrade (or improve) the underlying construct. In \Cref{sec:app_construct_validity}, we show that \metricname exhibits sensitivity, with mean scores falling by $6.9\%$, $9.0\%$, and $12.9\%$ when workflows are edited to remove AI-assisted steps under $5\%$, $10\%$, and $20\%$ perturbations, respectively (\Cref{fig:sensitivity_perturbations}). Stability checks that when perturbations are made that are not intended to affect the construct, the scores assigned remain consistent. We also find that \metricname is stable under perturbations where we do not expect reliance to change, including different \emph{reasoning effort}, different underlying models, and paraphrased workflow steps (\Cref{fig:stability_multiple_runs}).

\textbf{Criterion Validity.} Criterion validity measures the extent to which the proposed metric aligns with an external reference standard. In the absence of a single gold standard, we establish criterion validity for \metricname using the variation across experimental conditions in our user study (\Cref{sec:validation_user_study}), where prior work predicts higher reliance under time pressure.
\vspace{-2mm}
\section{Validating Our Measure of Reliance}
\label{sec:validation_user_study}
\vspace{-2mm}

\subsection{Study Design}
\label{sec:user_study_setup_design}

We conduct a between-subjects experiment to validate the proposed scalar measure of reliance, \metricname.
Users are recruited to complete coding tasks where the goal is to build functional web-based applications using any AI tools of their choice. We select four tasks spanning different skill domains, including graphics, database management, and asynchronous event handling.\footnote{We provide each set of task specifications in \Cref{sec:app_task_details}. Participants complete the tasks by satisfying these specifications.}

We introduce two experimental conditions that vary time constraints, a \shortcond condition ($1$ hour) and a \longcond condition ($4$ hours). 
Each task is completed by $5$ participants per condition, resulting in $n=20$ participants per condition. This design is motivated by robust evidence showing that time pressure increases reliance on external tools \citep{zakay1993impact, rice2009automation, jung2021towards, swaroop2024accuracy, haduong2024performance, rosbach2025automation, hua2025time}. Therefore, for \metricname to be a valid measure of reliance, it should assign higher scores on average in the \shortcond condition.
We evaluate this by comparing our measure across conditions using two-sample $t$-tests, and reporting the correlation with the binary condition label as an effect size.

\vspace{-2mm}
\subsection{Study Procedure and Participants}
\label{sec:user_study_procedure}
The study consists of three phases. First, participants complete a pre-task survey about their attitudes and typical usage of AI coding tools (\Cref{sec:app_pre_task_survey}). They then complete the coding task in the time frame corresponding to their assigned condition. This involves installing the \emph{workflow induction toolkit} \citep{wang2025ai}, and running the recording tool while finishing the task. 
The tool records all mouse and keyboard actions, while also taking screenshots when these are performed. Participants submit the recorded interaction data in addition to their final code via Google Drive or GitHub. Finally, participants complete a post-task survey consisting of: (a) a system recall component, where they answer $10$ questions about how task specifications were implemented in their code\footnote{We include system recall to capture how well users understand the systems they co-create, enabling us to interpret reliance in relation to task outcomes (\Cref{sec:normative_judgment}).}; (b) self-reported measurements on a Likert scale of (i) cognitive load, using the NASA-TLX scale, (ii) trust of AI output, (iii) perceived ownership of the final project, (iv) perceived distribution of cognitive work between the user and the tool. (c) participants' primary reasons for relying on the tool (\Cref{sec:app_post_task_survey}).
The study procedure was approved by an institutional ethics board (anonymized for peer review).

\begin{figure}
    \centering
    \includegraphics[width=\linewidth]{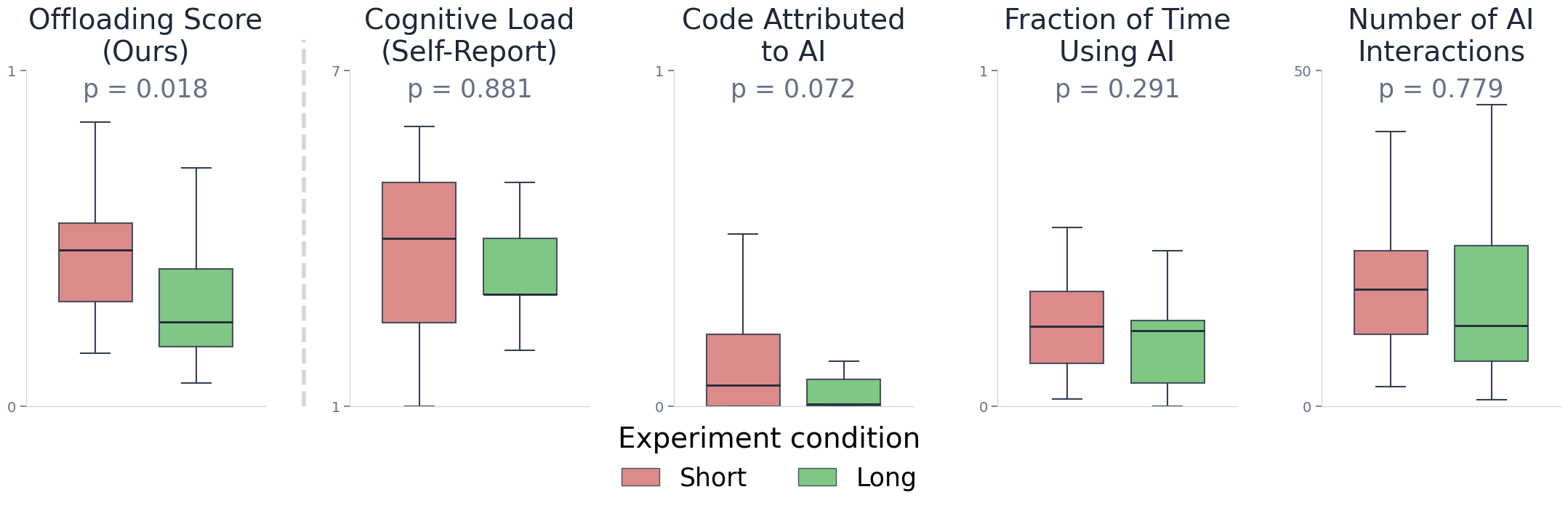}
    \caption{Comparison of different reliance measures across \shortcond and \longcond conditions in \Cref{sec:user_study_results}. \Metricname assigns significantly higher values for user reliance in the short condition (p = $0.018$), consistent with the expected trend from prior literature. In contrast, baseline measures based on AI usage (code attribution, time, interactions) and self-reported cognitive load have higher variance.
    \vspace{-2mm}}
    \label{fig:boxplot_comparison}
\end{figure}

\noindent
\textbf{Participants}
\label{sec:user_study_participants}
We recruit U.S.-based participants through the crowdworking platform Upwork. Participants were required to have completed at least three prior coding projects on the platform and to list AI coding tools they regularly use, ensuring familiarity with modern AI-assisted coding tools. We provide more details about participant recruitment  in \Cref{sec:app_participant_recruitment}. The final study consisted of $4$ tasks with $5$ participants per condition across $2$ experimental conditions ($40$ participants total).

\vspace{-2mm}
\subsection{Summary Statistics on Collected Traces}
\label{sec:user_study_summary_stats}
Workflows in \longcond condition are slightly longer on average than \shortcond ($176$ vs.\ $150$, $p>0.05$) with a lower proportion of AI-assisted steps ($16.40\%$ vs $9.68\%$, $p=0.03$) indicating more extended human involvement (\Cref{sec:app_summary_statistics}). 
Users use various tools from coding agents to chat assistants, with \texttt{Claude} and \texttt{ChatGPT} accounting for $70\%$ of workflows, followed by \texttt{Cursor} and \texttt{Gemini} (\Cref{fig:primary_tool_per_workflow}).\footnote{We calculate the primary tool in a workflow with a keyword matching process described in \Cref{sec:app_primary_tool}.} 

\vspace{-2mm}
\subsection{Calculating Measures of Reliance}
\textbf{Workflow Induction.}
We run the workflow induction tool on the collected interaction traces which results in a sequence of workflow steps that describes how the user completes the task with their tools. \Cref{sec:app_workflow_induction} provides more details of this process and a representative example (\Cref{sec:app_example_workflow}). 

\noindent
\textbf{Calculating Proposed Measures}
For each participant, we first identify the subset of workflow steps that involve interaction with the AI tool using \texttt{gpt-5.2}. We then compute \metricname for each participant by constructing human-only counterfactual workflows using \texttt{gpt-5.2}, following the procedure described in \Cref{sec:proposed_measure}.
We also use \texttt{gpt-5.2} to label the AI-assisted steps for \cogprocess and \outputuse labels according to the rubrics described in \Cref{sec:proposed_measure}. In addition to the step itself, we provide the surrounding workflow context and the overall task goal in the prompt (\Cref{sec:app_rubric_processes} and \Cref{sec:app_rubric_output}).
We report both group differences and correlations with the condition label for \metricname, and use the distributions over \cogprocess and \outputuse labels for descriptive analysis. 

\noindent
\textbf{Baseline measures.}
We also evaluate various baseline measures from the literature against the binary condition label. We report (a) total number of AI interactions, (b) lines of code generated by the tool(s) (\Cref{sec:app_code_attribution}), and (c) the fraction of time spent using the tool. We provide details on how we compute these measures in \Cref{sec:app_workflow_induction}.
In addition, we also include (d) first-person measurement of cognitive load using the NASA-TLX scale from the post-task survey. 

\vspace{-2mm}
\subsection{User Study Results}
\label{sec:user_study_results}

\noindent
\textbf{\Metricname reliably captures increased reliance under time pressure.}
From \Cref{fig:boxplot_comparison}, we observe that \metricname assigns higher reliance in the \shortcond condition ($p = 0.018$), as anticipated by the reference hypothesis for our user study. This provides evidence that \metricname captures variation in reliance induced by experimental manipulation. This effect also holds within each task individually, with a higher mean \metricname assigned to the \shortcond label on each task (\Cref{sec:app_reliance_by_task}). 

\noindent
\textbf{Comparison between \metricname and baseline measures of reliance.}
We also compare \metricname to baseline measures of reliance based on AI usage and self-reported experience. We expect a strong measure of reliance to be able to distinguish between users in our experimental conditions. From \Cref{fig:boxplot_comparison}, we observe that none of the baseline metrics show significant differences between conditions at the $5\%$ level, while \metricname is able to do so. We do note that the strongest usage-based baseline measure, the fraction of code attributed to the tool, approaches significance ($p = 0.072$). 
We also examine the correlation of each measure with the condition label in \Cref{fig:correlation_condition} and find that \metricname exhibits the strongest association (Pearson $r = 0.37$), compared to weaker correlations for baseline measures (\eg $r = 0.17$ for code attribution being the second strongest). Together, these results suggest that  \metricname provides a more sensitive measure of reliance compared to existing usage-based and self-reported measures.
We provide examples of \emph{points} of high reliance in workflows in \Cref{sec:app_example_high_reliance} where users rely on AI outputs with minimal iteration and use their tool to entirely design and implement a feature. 

\begin{figure}
    \centering
    \includegraphics[width=0.9\linewidth]{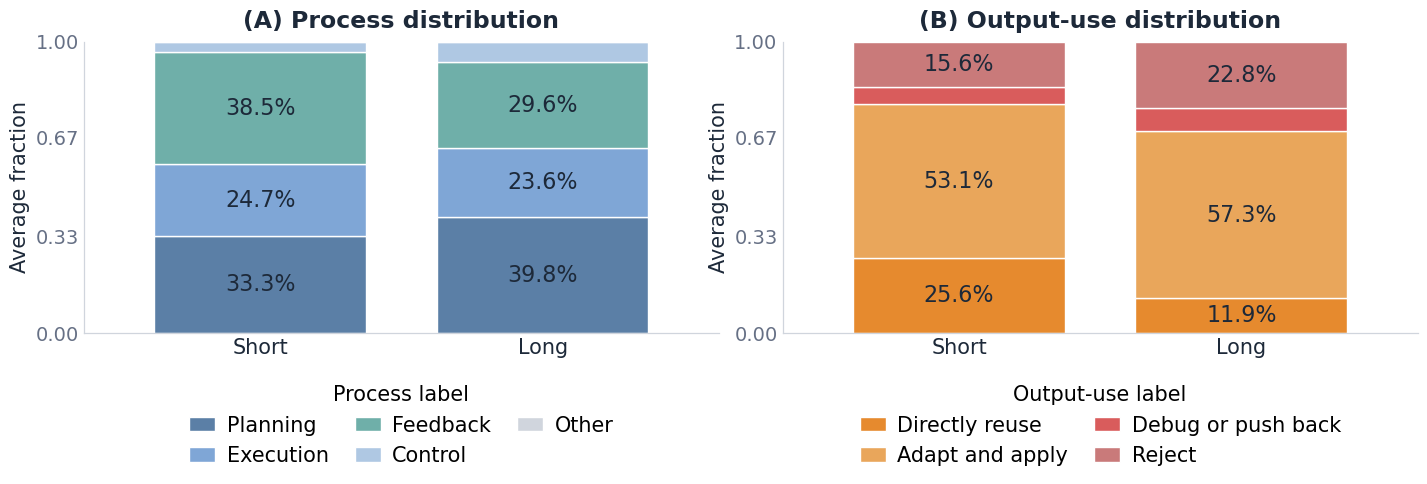}
    \caption{Distribution of \cogprocess and \outputuse labels across \shortcond and \longcond conditions, normalized by total number of labeled interactions. Users in the \shortcond condition directly execute subtasks with the tool 
    and reuse the model outputs, while users in the \longcond condition more frequently reject or adapt outputs, indicating more selective engagement. 
    \vspace{-3mm}}
    \label{fig:desc_label_dist}
\end{figure}

\noindent
\textbf{Higher reliance manifests as increased delegation and more direct reuse of outputs from the tool.}
As shown in \Cref{fig:desc_label_dist}, users in the \shortcond condition engage in more execution-oriented interactions ($38.5\%$ vs. $29.6\%$ in \longcond) and frequently directly reuse tool outputs ($25.6\%$ vs. $11.9\%$), suggesting that time pressure shifts users toward delegating task execution and incorporating outputs with minimal modification. In contrast, users in the \longcond condition reject outputs more often ($22.8\%$ vs. $15.6\%$ in \shortcond) and use the model more for planning-related processes ($39.8\%$ vs. $33.3\%$), indicating more selective engagement and greater user agency. We provide an illustrative example in \Cref{sec:app_example_pushback}.

\vspace{-1mm}
\section{Using \Metricname to Identify Conditions of (In)appropriate Reliance}
\label{sec:normative_judgment}
\vspace{-2mm}

\label{sec:normative_judgment_setup}
An important utility of a reliance metric is to help identify inappropriate reliance, such as determining whether a user’s reliance score exceeds a threshold to indicate overreliance. However, such normative judgments have to be made based on the user goal, or what constitutes undesirable outcomes of inappropriate reliance. 
Here we analyze how \metricname, in combination with target task outcome measures, can be used to identify when reliance may be considered (in)appropriate.

\noindent
\textbf{Code understanding as a desirable outcome for human-AI collaborative coding.}
\label{sec:normative_judgment_recall}
We consider maintaining a good understanding of the task (i.e., code being produced) as a desirable outcome. This choice is motivated by findings from software engineering literature on the risks of accumulating technical debt, including increased costs of future development and maintenance \citep{lim2012balancing, zazworka2013case, martini2015danger}. Since intuitively, relying on AI-assisted tools more likely creates risks instead of offering benefits for maintaining code understanding, here we only consider \textit{overreliance} (as opposed to underreliance) as the undesirable, inappropriate reliance.  Note that our approach can be applied to different normative goals. In practice, it is possible that the normative standards for different goals (e.g., what is the right reliance for harnessing productivity gain versus maintaining a good understanding) will differ, and one needs to make a priority judgment. 

\noindent
\textbf{Calculating user system recall as a proxy for code understanding.}
As a proxy for code understanding, we use \emph{system recall}, measured from the post-task survey (\Cref{sec:user_study_procedure}). This measure scores how well users understand the implementation details of the system they co-create with the tool. Each user answers $10$ questions about their system, which we evaluate using an LLM-as-judge against the reference implementation. Answers are scored on a scale as \emph{incorrect} (scored zero), \emph{partially correct} ($0.33$), \emph{mostly correct} ($0.67$), or \emph{fully correct} ($1.00$). We use \texttt{gpt-5.2} for performing this evaluation, and provide both additional details and the prompt used in \Cref{sec:app_system_recall_eval}. From these, we obtain an average code understanding score of the user about the system they submit. To illustrate defining appropriate reliance, we use a threshold of $0.33$ (\emph{partially correct}) as an indicator of ``good enough understanding''. In practice, such a threshold can be set empirically, by benchmarking the understanding level of programmers with good long-term productivity and agency. 

\vspace{-2mm}
\subsection{Identifying Conditions for Overreliance}
\label{sec:normative_judgment_results}
\begin{figure}
    \centering
    \includegraphics[width=0.75\linewidth]{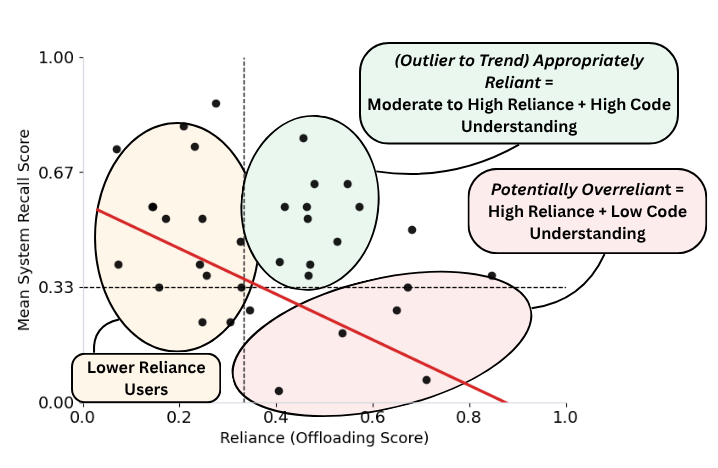}
    \caption{
    System recall ($y$-axis) vs. \metricname ($x$-axis) across participants. The red line shows a linear fit after excluding the moderate-to-high reliance, high-recall cluster in green, illustrating the stronger negative association between \metricname and system recall among the remaining users. The highlighted cluster suggests effective use of the tool while maintaining understanding (\Cref{sec:normative_judgment_results}).
    \vspace{-2mm}
    }
    \label{fig:behavior_patterns_partial}
\end{figure}

\noindent
\textbf{A general negative correlation between \metricname and system recall.} 
Intuitively, reliance should predict the cognitive risk of loss of understanding. We observe that the correlation of \metricname to system recall is higher ($-0.145$) than other baseline reliance measures (the strongest baseline is $-0.101$) (\Cref{fig:correlation_to_recall}). This offers more evidence of the validity of our metric. 
To dig deeper, in \Cref{fig:behavior_patterns_partial}, we plot reliance, measured by \metricname, versus system recall, measured by the average score obtained by the users on all questions.
When excluding users in the outlier cluster ($30\%$ of users in the green region in \Cref{fig:behavior_patterns_partial}, to be discussed below), we find that the Pearson correlation of \metricname and system recall increases to $-0.440$. For these users, we can define overreliance as the reliance level that leads to an undesirable level of system recall of below $0.33$, 
setting a threshold for overreliance as \metricname$=0.33$ as shown in \Cref{fig:behavior_patterns_partial}. 
Excluding the outlier clusters, points with higher \metricname generally fall below the $0.33$ of system recall, correctly identified to be overreliance. This predictive power allows us to set an operational threshold of overreliance for new interactions where the outcome of system understanding is not yet measured or observed.

\noindent
\textbf{Analyzing the outlier cluster with high \metricname and high system recall.} 
\Cref{fig:behavior_patterns_partial} indicates a cluster of points (green region) deviating from the linear fit, exhibiting high reliance and high recall. In their post-completion survey, $5$ (out of $11$) of them reported using the tool for coding features that they did not know how to implement by themselves. In a follow-up interview, one of these users noted ``not wanting to limit themselves to one line of thinking'' and hence involving the model in a back-and-forth planning loop. This distinct pattern suggests that 
the normative standard of appropriate reliance should be set differently for these users: when the tool is used for \textit{learning}, high reliance is often still appropriate for desirable outcomes. Our experiment does not provide enough data to identify the threshold for inappropriate reliance in this condition: the highest \metricname in this cluster is $0.65$, and the user still maintained a good enough understanding.
\vspace{-3mm}
\section{Conclusion}
\label{sec:conclusion}
\vspace{-2mm}
In this work, we introduce a process-oriented measure of reliance consisting of a scalar metric, \metricname, that quantifies the fraction of cognitive effort offloaded to AI tools using human counterfactual workflows, combined with more interpretable dimensions categorizing observed AI tool usage. Importantly, \metricname can be computed directly from interaction traces without requiring reference outputs, making it easily reusable across different tools and interfaces. Through a controlled user study, we show that \metricname captures variation in reliance aligned with known behavioral drivers, outperforming existing usage-based and self-reported measures. We further demonstrate that combining reliance with code understanding reveals distinct patterns of user behavior, showing the value in interpreting reliance in the context of task outcomes. We discuss limitations of our method in \Cref{sec:limitations}. Looking forward, as user modeling improves \citep{shaikh2026learning}, \metricname can be adapted to provide personalized estimates of reliance that account for individual differences in tool usage in the estimated counterfactuals, allowing users to reflect on their own behavior. \Metricname can also serve as a signal for training agents that actively mitigate overreliance by encouraging more balanced interaction patterns.

\section*{Acknowledgments}
We are thankful to Joachim Baumann, Hao Zhu, Ryan Louie, Omar Shaikh, Sunny Yu, Dora Zhao, Chenglei Si, Nishant Balepur, Nitish Joshi, Helena Vasconcelos, Judy Shen, and other members of the Stanford SALT Lab for their valuable feedback at various stages of the project. We thank the participants for the user study recruited from Upwork. This work was supported by an HAI grant, DSO lab, Open Philanthropy, Schmidt Sciences, a grant under the NSF CAREER IIS-2247357, ONR N00014-23-1-2420, and ONR N00014-24-1-2532.

\bibliographystyle{plainnat}
\bibliography{custom}

\appendix
\newpage

\section{Limitations}
\label{sec:limitations}

We note several limitations of our work. First, we validate \metricname against a single reference variable, time pressure, which, while grounded in prior literature, remains only one dimension of variation in reliance. Future work should evaluate the metric against a broader set of behavioral drivers and task settings. We note that our user study is limited to $40$ participants on programming tasks, and, while we do perform significance testing, may not generalize to other domains or populations.
The computation of \metricname depends on estimating human counterfactual workflows which may not always reflect realistic or optimal ways to achieve the same goal. Errors in this estimation directly affect the metric. The workflow induction from interaction traces is itself imperfect because the granularity of inferred steps may vary, and some steps may be misidentified or omitted which affects the results. More broadly, our formulation assumes that cognitive effort can be approximated through workflow steps, which may not capture all aspects that are not externally observable. We do note that as the research community invests more into workflow induction, user modeling and simulation, we expect the computation of \metricname to correspondingly improve as well. We also note a limitation that our formulation of \metricname does not consider \emph{personalized} counterfactual steps, instead opting for a general user counterfactual. We select this to ensure our measure is reusable and due to higher noise on estimating personalized workflows from limited interaction history. We acknowledge the large variance in user skill level and habits might mean that the general counterfactual is not equally applicable. An important line of future work is adapting \metricname to personalized workflow induction.

\section{Workflow Induction Process}
\label{sec:app_workflow_induction}

We broadly follow workflow induction as detailed in \citet{wang2025ai}. 
The input to the induction pipeline is the recording obtained from a participant session, including timestamped UI actions (clicks, key presses) and periodic screenshots. These raw traces are converted into a chronological action trajectory. First, these are segmented into contiguous work units using visual change between screenshots with mean squared error. These segments are then annotated with short natural-language descriptions of the visible activity using a VLM. The annotated segments are passed to an LLM, which groups consecutive segments into higher-level workflow activities. The final outputs are an induced natural language workflow and the merged activity segments, a time-aligned activity timeline with gaps/pauses inserted. 

We note a few hyperparameters from the process. The first is the threshold of similarity between screenshots for grouping. We set the MSE threshold to $500$, lower than the default, to ensure that no activities were lost. This resulted in a few \emph{empty} workflow steps, examples include ``No meaningful action visible'', and ``Idle or non-captured actions with no significant changes''. We manually filtered these out through a series of rule-based approaches prior to analysis. The second is the model used for induction. We use \texttt{gpt-5.1} for workflow induction accessed via the API, providing screenshots as images when needed. The prompts for the workflow induction remain the same as the released toolkit. 
We note that advances in the workflow induction pipeline should transfer to more reliable measurement for \metricname. We share the code for our analysis at \href{https://github.com/vishakhpk/offloading-score}{this linked repository}.

\begin{figure}
    \centering
    \includegraphics[width=0.475\linewidth]{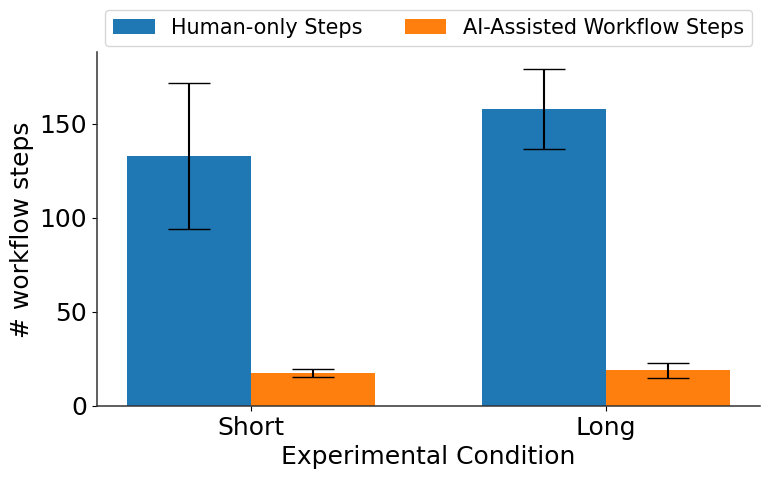}
    \caption{Average number of steps in each workflow, divided into AI-assisted and human-only steps. We observe slightly longer trajectories in the \longcond condition, with slightly higher human-only steps in \longcond with similar AI-assisted steps leading to slightly lower proportion of AI-assisted steps. }
    \label{fig:workflow_step_stats}
\end{figure}%

\subsection{Summary statistics of workflows}
\label{sec:app_summary_statistics}
We report summary statistics of the collected workflows in \Cref{fig:workflow_step_stats}. The total number of steps in workflows from the \longcond condition is slightly longer on average than \shortcond ($176$ long vs $150$ short, $p>0.05$ on a two-sided Welch test), with increases in human-only steps ($157$ long vs $132$ short) and almost identical AI-assisted steps ($18$ long vs $17$ short), indicating more extended human involvement in the trajectories (\Cref{fig:workflow_step_stats}). As a result, the relative proportion of AI-assisted steps is higher for the short condition ($9.68\%$ long vs $16.40\%$ short, $p=0.03$ on a two-sided Welch test). 

\subsection{Primary tool in each workflow}
\label{sec:app_primary_tool}
\begin{figure}
    \centering
    ~ 
    \includegraphics[width=0.65\linewidth]{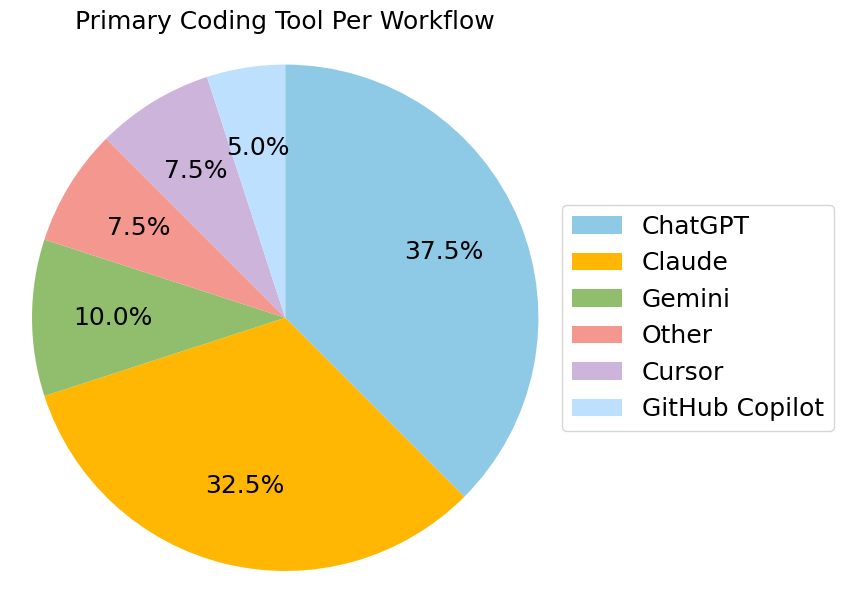}
    \caption{Primary tool used in each workflow, calculated by mentions in the induction process.}
    \label{fig:primary_tool_per_workflow}
\end{figure}

\begin{figure}
    \centering
    \includegraphics[width=0.75\linewidth]{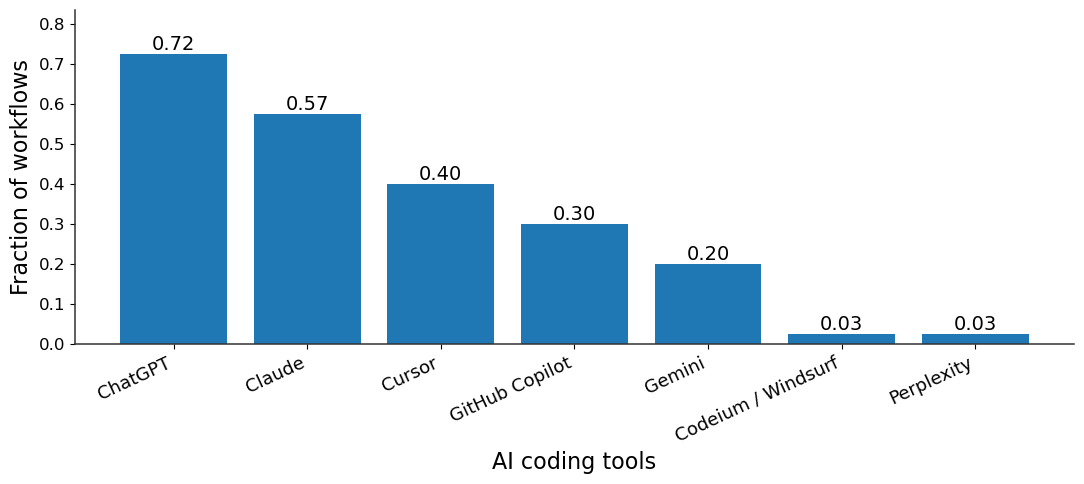}
    \caption{Fraction of workflows in which there is a keyword mention associated with different AI-coding tools.}
    \label{fig:all_tool_mentions}
\end{figure}

Given the sequence of workflow steps in natural language, we estimate the tools used in each with a simple keyword-based search. We provide the list of keywords used to map to each different tool at the end of this section. We count the frequency of each tool mentioned in each workflow, and plot the fraction of workflows in which each tool is mentioned in \Cref{fig:all_tool_mentions}. Note that these do not sum up to $100\%$ as users can use multiple tools in the same workflow. 
In \Cref{fig:primary_tool_per_workflow}, we instead report the `primary' tool used in each workflow or the tool with the most mentions. 

\begin{verbatim}
assistant_keywords = {
    "ChatGPT": [
        "chatgpt", "gpt-4", "gpt4", "gpt 4", "gpt-4o", 
        "gpt 4o", "gpt-3.5", "gpt 3.5", "openai", 
        "o1", "o3", "codex"
    ],
    "GitHub Copilot": [
        "github copilot", "copilot"
    ],
    "Claude": [
        "claude", "anthropic", "claude sonnet", 
        "claude opus", "claude haiku"
    ],
    "Gemini": [
        "gemini", "bard", "google ai studio"
    ],
    "Cursor": [
        "cursor"
    ],
    "Codeium / Windsurf": [
        "codeium", "windsurf"
    ],
    "Tabnine": [
        "tabnine"
    ],
    "Replit Ghostwriter": [
        "replit ghostwriter", "ghostwriter", "replit ai"
    ],
    "Perplexity": [
        "perplexity"
    ],
}
\end{verbatim}

\subsection{Example Workflow}
\label{sec:app_example_workflow}
Here we provide an example output of workflow induction for illustrative purposes. 
\begin{verbatim}
- (reading generation) Send prompt and review AI’s step-by-step UI/UX 
improvement plan and to-do list.
- (writing code) Modify CSS styles while following AI’s implementation
order for visual improvements.

...

- (reading generation) Scroll AI chat to review “Step 1: Visual 
foundation” and HTML restructure.; Click “Review” to inspect proposed
HTML/CSS changes for index.html.
- (test manually checking) Start timer and check localStorage in 
DevTools Application tab.
[598-598] (reading code) Review CSS rules in index.html by clicking
editor area.

...

- (reading generation) Review AI “Modern minimal CSS design” guidance
while editing CSS.
- (writing prompt) Send prompt asking AI to refactor layout for 
improved alignment and spacing.
- (test manually checking) Start timer and observe automatic state 
change from Break to Work.
- (writing code) Review CSS layout changes in VS Code and then 
visually check in browser.
- (reading generation) Click “Review” in AI pane to inspect/apply
layout modifications.
\end{verbatim}

\section{Implementation Details For \Cref{sec:formulation}}
\label{sec:app_implementation_details}

\subsection{Rubric for tasks offloaded to AI}
\label{sec:app_rubric_processes}

\begin{verbatim}
ANNOTATION_PROMPT = dedent(
    """
You are an expert cognitive analyst. You will analyze a sequence of 
actions taken by two agents in a collaborative coding setting.

Your task is to classify **one target action** into a cognitive process 
category using the theoretical framework below.

---

## INPUT FORMAT

You are provided with:

Previous Action: <text describing what happened before the target action>
**Current Action**: <text describing the action to be labeled>
Next Action: <text describing what happens after the target action>

Broader Workflow Context:
- Previous Workflow Step: <workflow step before the matched step, if any>
- Matched Workflow Step: <matched workflow step>
- Next Workflow Step: <workflow step after the matched step, if any>

You must classify the **current_action** based on its function, using both
 the local action context and workflow context.

---

## OUTPUT FORMAT

Return a single JSON object with the following keys:

{
  "process_type": "planning | execution | feedback | control",
  "justification": "<brief reasoning for the process type, referencing 
  the surrounding action or workflow context when relevant>"
}

---

## THEORETICAL FRAMEWORK

This scheme adapts the **Flower & Hayes (1981) Cognitive Process Model** for 
programming workflows.
Cognitive processes are recursive and context-dependent — planning, 
executing, and evaluating occur repeatedly throughout a coding session.

### 1. Process Categories

**PLANNING**
Formulating or revising goals, structuring code, or selecting solution strategies.
→ Examples: outlining function design, deciding on data structures, sketching 
pseudocode, reading task description before coding.

**EXECUTION**
Translating plans into concrete code or commands.
→ Examples: typing new functions, editing syntax, refactoring code, 
implementing logic, running code snippets.

**FEEDBACK / MONITORING**
Evaluating outputs, debugging, interpreting logs or test results, comparing 
alternatives.
→ Examples: reading error messages, analyzing stack traces, inspecting 
model output, consulting docs to validate reasoning.

**CONTROL / COORDINATION**
Meta-level actions that regulate task flow or transitions between phases.
→ Examples: pausing to reprioritize subtasks, switching from exploration to 
implementation, deferring a fix, or prompting the model for clarification.

---

## DECISION RULES

1. Use previous and next action context to interpret what cognitive function
 the current action serves.
2. Use the workflow context to disambiguate broad or underspecified actions.
3. Label actions by the function they perform, not by vague intention.
   - E.g., suggesting code = planning, executing code = execution, 
   showing errors = feedback.

---

Previous Action: {previous_step_text}
Current Action: {current_step_text}
Next Action: {next_step_text}
Previous Workflow Step: {previous_workflow_step}
Matched Workflow Step: {matched_workflow_step}
Next Workflow Step: {next_workflow_step}
"""
)
\end{verbatim}

\subsection{Rubric for engaging with AI output}
\label{sec:app_rubric_output}

\begin{verbatim}
OUTPUT_USE_NEXT_STEPS_PROMPT = dedent(
    """
You are an expert analyst studying how users integrate AI-generated 
content into their workflow.

Your task is to classify how the AI-assisted step at a specific moment is 
used in the subsequent workflow.

==================================================
INPUT DESCRIPTION
==================================================

You will receive the following inputs:

TASK_DESCRIPTION
- The overall task the user is trying to accomplish.
- This provides global context for understanding how the AI-assisted step
 might contribute.

PREVIOUS_HUMAN_STEP
- The nearest preceding human-authored step before the AI-assisted step
 being analyzed.
- This approximates what the user was doing or asking for around this moment.

CURRENT_AI_ASSISTED_STEP
- The current AI-assisted step being analyzed.
- This is the content whose downstream use you must evaluate.

WORKFLOW_CONTEXT
- The matched workflow step for the AI-assisted step, plus one workflow step
 before and one after.
- This is only secondary context and may be imperfect.
- Use it only as light background framing, not as primary evidence.

NEXT_STEPS
- A sequence of subsequent actions (from the user and/or tool) that occur after
 the AI output.
- These steps provide the primary behavioral evidence of how (or whether) the
 AI output was used.

Your classification must be based ONLY on observable evidence in 
NEXT_STEPS, while using TASK_DESCRIPTION and WORKFLOW_CONTEXT
 as contextual framing.
If NEXT_STEPS and WORKFLOW_CONTEXT seem to point in different 
directions, trust NEXT_STEPS.

==================================================
INPUT
==================================================

TASK_DESCRIPTION:
<<<
{task_description}
>>>

PREVIOUS_HUMAN_STEP:
<<<
{previous_human_step}
>>>

CURRENT_AI_ASSISTED_STEP:
<<<
{current_ai_assisted_step}
>>>

WORKFLOW_CONTEXT:
<<<
Previous Workflow Step: {previous_workflow_step}
Matched Workflow Step: {matched_workflow_step}
Next Workflow Step: {next_workflow_step}
>>>

NEXT_STEPS:
<<<
{next_steps_sequence}
>>>

==================================================

Your goal is to determine how the CURRENT_AI_ASSISTED_STEP is integrated
 into the workflow based on evidence in NEXT_STEPS.

--------------------------------------------------
CLASSIFICATION LABELS
--------------------------------------------------

Choose exactly ONE label.

- Reuse
The user directly reuses the AI-assisted content with minimal or trivial 
changes.
Examples:
- copying or pasting AI-generated text, code, or commands
- executing a command clearly provided by the AI
- accepting or applying an AI suggestion with only minor visible modification
- incorporating AI-generated wording or code into docs/files with little 
transformation

- Apply
The user takes an idea, method, structure, or recommendation from the 
AI-assisted content and adapts it to their own context in a visible way.
Examples:
- making a concrete edit, configuration change, test, or revision that clearly 
follows the AI's recommendation
- reimplementing an AI-suggested idea in a different form
- using AI recommendations as a basis for edits, tests, prompts, or plan 
changes that are clearly connected to that specific guidance but not direct
 copy-paste
- following the AI's suggested approach, but with noticeable adaptation 
rather than direct reuse

- Pushback
The user tests, questions, challenges, corrects, or follows up on the 
AI-assisted content because it appears insufficient, problematic, incomplete,
 or still unresolved.
Examples:
- identifying that the AI suggestion is wrong, incomplete, mismatched, or 
does not fully solve the problem
- asking follow-up questions, debugging, or testing because the AI output
 appears flawed or insufficient
- revising direction or seeking correction because the AI content was 
evaluated and found wanting

- Reject
There is no clear observable evidence that the AI-assisted content was 
meaningfully used in subsequent steps.
Examples:
- later steps are unrelated or only loosely related
- the user continues the task, but there is no visible trace that this 
specific AI-assisted step influenced those actions
- the connection is only thematic or speculative

--------------------------------------------------
DECISION GUIDANCE
--------------------------------------------------

Use a balanced standard:

- Do not require perfect proof for Apply. But Apply should only be used 
when NEXT_STEPS show a concrete downstream action that is 
specifically responsive to the AI-assisted content.
- Do not assign Reuse unless the connection is fairly direct.
- Prefer Reuse over Apply when the later steps show direct uptake of
 the AI content with little transformation, even if the user also verifies
  or lightly adjusts it afterward.
- Do not assign Reject merely because the later steps are not 
copy-paste explicit.
- Prefer Reject over Apply when the later steps are only generic 
continuation of the same task, such as browsing nearby files, generic
 testing, or continuing implementation without a clear visible trace of 
 this specific AI-assisted step.
- Prefer Apply over Reject only when there is a specific visible connection
 between the AI-assisted content and what the user does next, such as 
 implementing the recommended change, testing the suggested fix, or 
 revising a file or plan in a way that clearly matches the guidance.
- Prefer Pushback when the user visibly responds to the AI output by 
probing, challenging, debugging, or seeking correction because the 
output seems wrong, incomplete, mismatched, or unresolved.
- Do not use Pushback for ordinary implementation or generic 
verification. Use it only when the later behavior indicates friction with 
the AI output itself.
- Prefer Reject when the relationship is only thematic, vague, or 
speculative.
- Prefer NEXT_STEPS over WORKFLOW_CONTEXT if they conflict.
- Similar topic alone is not enough; there should be some visible 
downstream trace.

--------------------------------------------------
IMPORTANT RULES
--------------------------------------------------

1. Base your decision ONLY on observable evidence in NEXT_STEPS.
2. Do NOT infer hidden learning unless behaviorally reflected.
3. Do NOT use a staged or linear interpretation. Choose the single 
best-fitting label from the set above.
4. WORKFLOW_CONTEXT is secondary and may be imperfect.
5. Running tests, editing files, reading docs, or continuing the task 
support Apply only when they clearly instantiate or verify the AI's specific
 recommendation, not when they are merely generic continuation.
6. Return exactly one classification.

--------------------------------------------------
OUTPUT FORMAT
--------------------------------------------------

Return a JSON object in this exact format:

{
  "label": "<Reuse | Apply | Pushback | Reject>",
  "justification": "<brief explanation referencing concrete evidence 
  from NEXT_STEPS>"
}

Return ONLY the JSON object.
No additional commentary.
"""
)
\end{verbatim}

\subsection{Rubric for creating human-only counterfactual}
\label{sec:app_human_counterfactual}

\begin{verbatim}
You are rewriting a single step in a human-AI collaborative 
coding action with a realistic user-only steps. Essentially, you 
are estimating how a human would have accomplished the same
step without any AI assistance, while matching the granularity of
surrounding user actions. You can introduce any additional tools
the user might have used to accomplish the task in the absence 
of AI. Your goal is to replace the single human-AI step with an 
appropriate number of user-only steps that accomplish the same 
code changes or actions.

GOAL:
Replace all AI actions that occur in the current TOOL turn with 
an appropriate number of USER ONLY turns that:
- Accomplish the same code changes or actions performed 
that the human-AI step did
- Match the granularity of surrounding workflow steps
- Are realistic human workflow actions
- Do NOT include any AI actions or references to AI assistance
- Introduce any additional tools they might have used to 
accomplish the task in the absence of AI
- Are NOT too small (no single keystrokes)
- Are NOT too large (no "wrote entire file")

CONTEXT — PREVIOUS TURNS:
{context_block}

TOOL TURN TO REWRITE:
{tool_block}

NEXT TURN (for alignment):
{next_block}

OUTPUT FORMAT:
- Return ONLY the reconstructed steps, a list of strings which 
are the steps that would replace
  the single human-AI step if the user did not have access to 
  any AI tools.
- Do NOT include explanations.
- Do NOT include markdown.
"""
 
\end{verbatim}

\section{Metric Validity.}
\label{sec:app_metric_validity}
\paragraph{Criterion Validity}
Criterion validity measures the extent to which the proposed metric aligns with an external reference standard. In the absence of a single gold standard, we establish criterion validity for \metricname using the variation across experimental conditions in our user study (\Cref{sec:validation_user_study}), where prior work predicts higher reliance under time pressure. 

\paragraph{Content Validity}
Content validity assesses whether a metric captures the construct it is intended to measure. For constructs that are difficult to quantify directly, we follow prior work on automatic metric design \citep{ryan2025autometrics} by providing transparency into the metric computation and supporting evidence about the different components of our measurement. 
\metricname naturally affords transparency as the score is computed via a simple fraction of observed and counterfactual workflow steps. Users can inspect their workflow and the mapping of each AI-assisted step to corresponding human-only counterfactual steps to check whether the measurement reflects their behavior. In our user study, we validate the plausibility of these counterfactuals through participant responses (\Cref{sec:user_study_results}). We further support this by showing that the generated counterfactuals align with human workflows on the same tasks with significance at the $5\%$ level on a permutation test using a dataset of human and agent workflows from \citet{wang2025ai}.
We also verify the LLM-as-judge annotations of descriptive dimensions with human judgments (\Cref{sec:app_annotation_llm_as_judge}).

\paragraph{Construct Validity}
Construct validity asks whether a metric behaves consistently with the underlying construct it is intended to measure. Following \citet{campbell1959convergent}, we operationalize construct validity as robustness, evaluated through sensitivity and stability. Sensitivity tests whether a metric assigns systematically lower (or higher) scores when examples are perturbed in a manner meant to degrade (or improve) the underlying construct. Stability checks that when perturbations are made which are not intended to affect the construct, the scores assigned remain consistent. In \Cref{sec:app_construct_validity}, we verify that \metricname exhibits (a) sensitivity by showing that manually editing workflows to remove AI-assisted steps reduces metric scores (\Cref{fig:sensitivity_perturbations}); and (b) stability by showing that metric scores remain consistent to perturbations where the human counterfactual is generated in ways where we do not expect the reliance to change (\eg different \emph{reasoning effort}, different underlying model, paraphrasing steps) (\Cref{fig:stability_multiple_runs}).

\subsection{Validation of human-only counterfactual on prior dataset}
\label{sec:app_validate_human_counterfactual}
\paragraph{Dataset and our usage} To provide a data-driven validation of the procedure for creating human-only counterfactual workflows (\Cref{sec:proposed_measure}), we use the dataset collected as part of \citet{wang2025ai} that contains recorded workflows for four short-horizon software engineering tasks. These short-horizon tasks are useful for our setting because they resemble the kinds of steps that users offload to tools within longer coding workflows, including those in our user study. For each task, the dataset includes multiple human workflow traces, as well as AI-agent trajectories. In this validation, we use only the task instructions, or the request to the tools, and our prompting procedure from \Cref{sec:proposed_measure} to generate synthetic human-only counterfactual workflows, and we compare these against the recorded human workflows for the same tasks.
For each task, we create $5$ synthetic counterfactual workflows. We then embed each of these human and counterfactual workflows using a SentenceTransformer model (\texttt{sentence-transformers/all-MiniLM-L6-v2}). 
We hypothesize that workflows from the same task should cluster together, so for each synthetic workflow, we compute a statistic equal to its average similarity to human workflows from the same task minus the average similarity to human workflows from the other tasks.

\paragraph{Significance testing for median in-cluster similarity}
We first perform a one-sided Wilcoxon signed-rank test on these per-workflow statistics against the null hypothesis that their median is not greater than zero. We find strong evidence that the statistics are positive ($p \approx 1e^{-6}$), indicating that synthetic workflows are substantially more similar to human workflows from the same task than to those from other tasks.

\paragraph{Permutation test for cluster assignment}
We also perform a one-sided permutation test on the same statistic. For each synthetic workflow, we first compute the difference between its average similarity to human workflows from the same task and its average similarity to human workflows from other tasks. We then take the mean of these differences across all synthetic workflows. To form a null distribution, we repeatedly flip the sign of each workflow-level difference at random and recompute the mean, using $10{,}000$ random sign assignments. This null distribution represents the values of the test statistic that would be expected if counterfactual workflows were no more similar to human workflows from the same task than to those from other tasks. The p-value is the fraction of permuted means that are at least as large as the observed statistic.
This test also yields a highly significant result ($p < 10^{-4}$).

\paragraph{Takeaway.} These results show that the synthetic counterfactual workflows are semantically closer to the corresponding human workflows than to workflows from other tasks. While this does not establish that the synthetic workflows are exact reconstructions of true human-only behavior, it provides evidence that our procedure produces reasonable task-aligned approximations. We also verify our approximations with participants in our user study in \Cref{sec:user_study_results}.

\begin{figure}
    \centering
    \includegraphics[width=\linewidth]{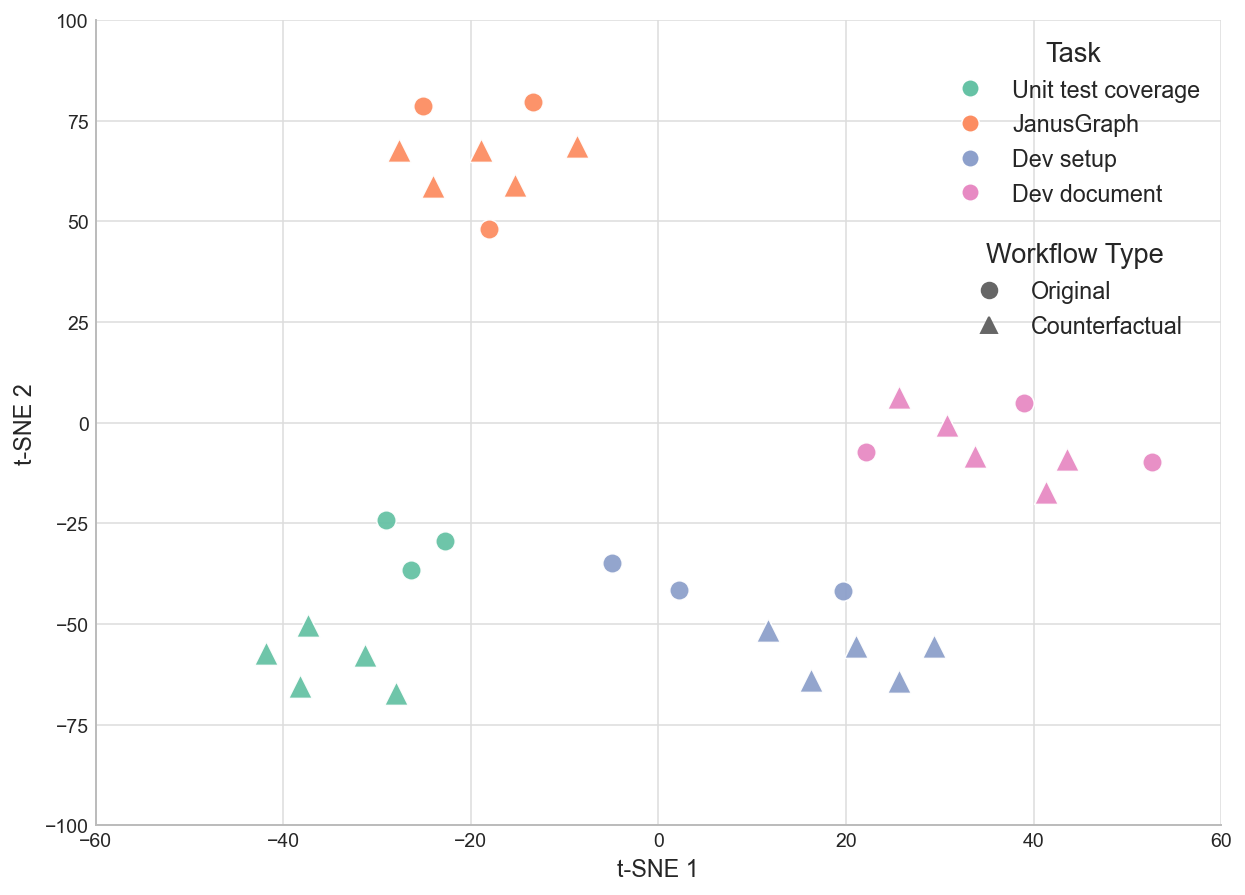}
    \caption{t-SNE visualization of workflow embeddings for recorded human workflows and synthetic counterfactual workflows across the four software engineering tasks (\Cref{sec:app_validate_human_counterfactual}). Colors indicate different tasks, and the marker shape indicates human versus synthetic workflows.}
    \label{fig:clustering_workflows}
\end{figure}

\paragraph{Visualization of clusters in 2D.} We also visualize the human and synthetic workflows in two dimensions using t-SNE applied to the workflow embeddings (\Cref{fig:clustering_workflows}).

\subsection{Verifying the quality of human counterfactuals with human annotations}
\label{sec:app_human_validation_counterfactual}
\begin{wrapfigure}{r}{0.4\linewidth}
    \centering
    \includegraphics[width=\linewidth]{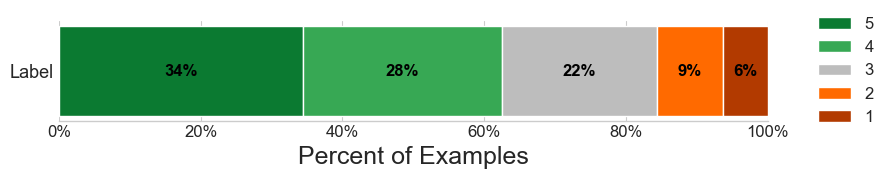}
    \caption{Distribution of Likert-scale ratings for generated human counterfactual steps, showing that most are perceived as plausible alternatives (\Cref{sec:user_study_results}).}
    \label{fig:likert_counterfactuals}
\end{wrapfigure}
The computation of \metricname relies on estimating human counterfactual workflows (\Cref{sec:pre_validation_metrics_validity}). We additionally validate the reliability of the counterfactuals with human annotations. We ask $20$ randomly sampled users to evaluate $5$ counterfactual steps from \emph{their own} workflows. For each step, users rate whether the proposed human-only alternative is a reasonable way to achieve the same sub-goal on a Likert scale from $1$ (Strongly Disagree) to $5$ (Strongly Agree). From \Cref{fig:likert_counterfactuals}, over $60\%$ of steps receive ratings of $4$ or $5$, while only $15\%$ are rated in disagreement ($1$ or $2$), indicating that the generated counterfactuals are generally perceived as plausible by participants.\footnote{We note that as the research community increasingly invests in user modeling \citep{shaikh2025creating, shaikh2026learning} and simulation \citep{park2023generative, park2024generative}, the computation of \metricname will continue to improve. We discuss this further in \Cref{sec:conclusion}.}

\subsection{Establishing construct validity through sensitivity and stability of \metricname}
\label{sec:app_construct_validity}

\begin{figure}
    \centering
    \includegraphics[width=0.6\linewidth]{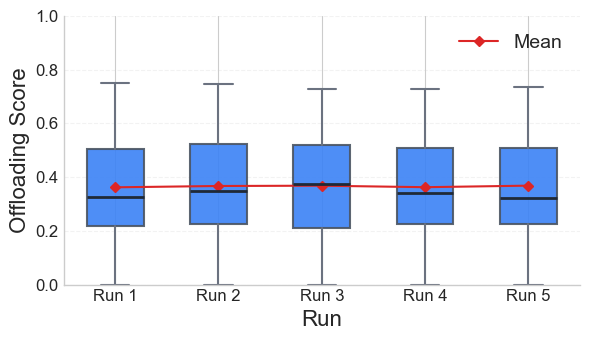}
    \caption{Variation in the distribution of \metricname scores across $5$ runs described in \Cref{sec:app_construct_validity}. We find that \metricname remains stable under these perturbations.}
    \label{fig:stability_multiple_runs}
\end{figure}

\paragraph{Stability}
Stability measures the ability of a metric to assign consistent scores under perturbations for which the quantity being measured, here \metricname, is not truly affected. \Cref{fig:stability_multiple_runs} shows the distribution of scores assigned on $5$ runs where, (a) Runs $1$ and $2$ are identical inputs where we re-run the counterfactual estimation with two different reasoning effort settings with the \texttt{gpt-5.2}; (b) Run $3$ varies the underlying model to \texttt{gpt-5-mini}; (c) Runs $4$ and $5$ was where we paraphrase a random sample of $20\%$ steps in the workflow with \texttt{gpt-5-mini} before rerunning the \metricname calculation with \texttt{gpt-5.2-mini}. We observe minimal differences across the runs, both in the spread of scores as well as aggregate measures (mean, median, quartiles). This confirms that the pipeline for calculating \metricname is stable.

\begin{figure}
    \centering
    \includegraphics[width=0.75\linewidth]{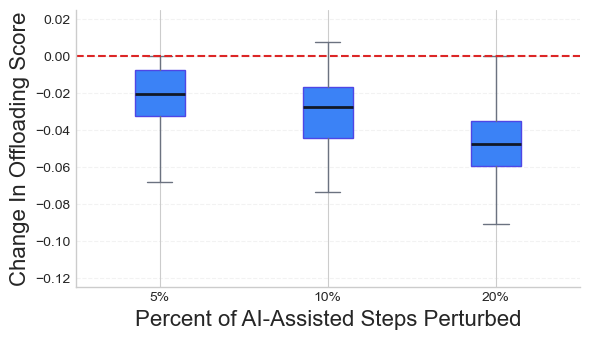}
    \caption{Variation in the distribution of \metricname values when perturbing $5\%$, $10\%$, and $20\%$ of AI-assisted steps in a manner designed to reduce reliance as described in \Cref{sec:app_construct_validity}. We find that \metricname is sensitive to each of these perturbations with significance, and larger perturbations lead to more decrease in \metricname scores.}
    \label{fig:sensitivity_perturbations}
\end{figure}

\paragraph{Sensitivity} Sensitivity measures the ability of a metric to assign lower scores under perturbations that are intended to produce lower reliance. 
To obtain targeted perturbations where users are less reliant, we randomly select $5\%$, $10\%$, and $20\%$ of AI-assisted steps from the collected workflows and replace these with the steps where the user did not use the tool, i.e., the human counterfactuals. From \Cref{fig:sensitivity_perturbations}, we see that perturbing workflows to reduce AI usage led to a systematic decrease in \metricname. Mean values fell in the $5\%$, $10\%$, and $20\%$ replacement conditions by $6.9\%$, $9.0\%$, and $12.9\%$. These decreases were significant at the $5\%$ level under both paired $t$-tests and Wilcoxon signed-rank tests for all three levels, indicating that the metric is sensitive to true reductions in AI usage.

\begin{figure}
  \centering
  \includegraphics[width=0.6\linewidth]{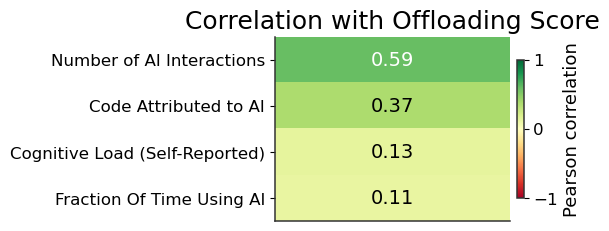}
  \caption{Pearson correlation scores between different baseline measures of reliance and \metricname. \metricname is strongly correlated with usage-based baseline measures.}
  \label{fig:correlation_baseline_measure}
\end{figure}

\begin{figure}
    \centering
    \includegraphics[width=0.6\linewidth]{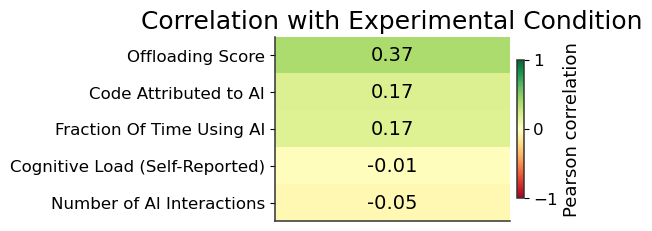}
    \caption{Pearson correlation scores between various measures of reliance and the experimental condition label. \Metricname correlates most strongly with the label.}
    \label{fig:correlation_condition}
\end{figure}%

\begin{figure}
    \centering
    \includegraphics[width=0.6\linewidth]{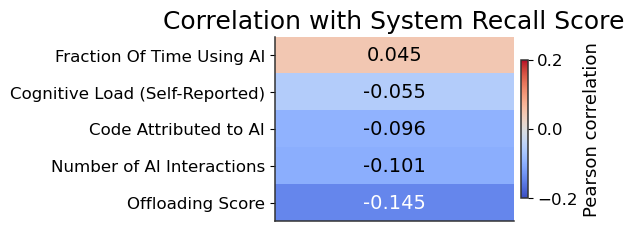}
    \caption{Pearson correlation between different measures of reliance, and task outcome, measured as a binary $\{0, 1\}$ value based on whether the user's system recall score is $\geq 0.33$ (\Cref{sec:normative_judgment}).}
    \label{fig:correlation_to_recall}
\end{figure}


\subsection{Validating LLM-as-judge with human annotations}
\label{sec:app_annotation_llm_as_judge}
To validate the use of LLM-as-judge for labeling workflow steps, we collect human annotations for the cognitive process and output-use labels introduced in \Cref{sec:formulation}. We recruit $6$ professional programmers from Upwork as annotators. Each example is annotated independently by two different annotators. For each labeling task, we sample $100$ workflow steps from different users and provide annotators with local context (the preceding and following three steps) to support consistent labeling. 
We first evaluate inter-annotator agreement to assess the reliability of the labeling scheme. For cognitive process labels, we observe $78\%$ agreement with Cohen’s $\kappa = 0.803$. For output-use labels, agreement is $84\%$ with $\kappa = 0.854$, indicating substantial agreement for both tasks.
We then evaluate LLM-as-judge performance by comparing model predictions against the majority human label. We consider three models: \texttt{gpt-5-mini}, \texttt{gpt-5.2}, and \texttt{gpt-5}. As shown in \Cref{tab:iaa_llm}, \texttt{gpt-5.2} achieves the highest agreement with human annotations across both tasks (80\% for cognitive process and 81\% for output use), and we use it for all subsequent analyses.

\begin{table}[t]
\centering
\small
\begin{tabular}{lcccccc}
\toprule
Task & Categories & Human Agreement & Cohen’s $\kappa$ & \texttt{gpt-5-mini} & \texttt{gpt-5.2} & \texttt{gpt-5} \\
\midrule
Cognitive process & $4$ & $78\%$ & $0.803$ & $63\%$ & $80\%$ & $74\%$ \\
Output use & $4$ & $84\%$ & $0.854$ & $75\%$ & $81\%$ & $76\%$ \\
\bottomrule
\end{tabular}
\vspace{4pt}
\caption{Human inter-annotator agreement and LLM-as-judge accuracy (against majority human labels) for cognitive process and output-use annotations. Metrics for each task are reported on $100$ randomly sampled examples, and $2$ human annotators scored each example.} 
\label{tab:iaa_llm}
\end{table}

\section{User Study Details}
\label{sec:app_user_study_details}

\subsection{Participant Recruitment}
\label{sec:app_participant_recruitment}
We recruit U.S.-based participants through the crowdworking platform Upwork.\footnote{\href{https://www.upwork.com/}{Upwork} is an online platform where freelancers sign up to complete various tasks.} When evaluating respondents to the job posting, we filtered for participants who had completed at least three prior coding projects on the platform and were able to list coding tools they regularly use, ensuring familiarity with contemporary AI-assisted coding tools. Participants were compensated at a rate of $\$20$ to $\$30$ per hour according to their self-proposed hourly rate. 
In total, we recruited $44$ participants and randomly assigned them to one of the tasks and to one of the experimental conditions. We excluded $4$ participants due to anomalous completion behavior (e.g., finishing in under 10 minutes, taking more than 6 hours, or not using any AI tools), and replaced them to maintain balance. The final study consisted of $4$ tasks with $5$ participants per condition across $2$ experimental conditions ($40$ participants total)

\subsection{Pre-Task Survey}
\label{sec:app_pre_task_survey}
\paragraph{Overview}
We collect information about participants' prior experience with AI-assisted coding tools, including which tools they use, their level of access, typical use cases within their workflow, and the extent to which AI contributes to code generation. 

\paragraph{Survey Questions}
\begin{enumerate}
    \item \textbf{What is your Upwork Username} (required)
    
    \item \textbf{AI coding tools that you have used in the last 30 days} (check all that apply)
    \begin{itemize}
        \item Github Copilot
        \item Cursor
        \item Claude Code
        \item Codex
        \item Manus
        \item CodeWhisperer
        \item Tabnine
        \item Autocomplete from your IDE
        \item Custom or in-house AI tools
        \item Other
    \end{itemize}
    
    \item \textbf{What tier of access do you use for these services} (select one)
    \begin{itemize}
        \item Free versions only
        \item Paid subscription
    \end{itemize}
    
    \item \textbf{Typically what do you use AI for in your workflow?} (check all that apply)
    \begin{itemize}
        \item Writing new modules with AI agents
        \item Autocomplete / inline suggestions
        \item Debugging errors
        \item Explaining unfamiliar code
        \item Refactoring / cleanup
        \item Writing tests
        \item Documentation / comments
        \item Code review assistance
        \item Searching APIs / libraries
        \item Other
    \end{itemize}
    
    \item \textbf{Lines of code written by AI in a typical coding project} (select one)
    \begin{itemize}
        \item $> 75\%$
        \item Between $25\%$ and $75\%$
        \item Less than $25\%$
        \item I don't use AI to write lines of code, only for other things
    \end{itemize}
\end{enumerate}

\subsection{Task Details}
\label{sec:app_task_details}

Participants complete one of four simple web development tasks designed to capture realistic programming workflows. The goal of each task is to build a self-contained local web application while satisfying a list of mandatory requirements provided. Here we list the four tasks.

\subsubsection{Task 1: Mindful Break Timer}
Build a simple local web app that lets users run focused work timers and receive mindful break prompts between sessions. The goal is a clean, usable timer that encourages healthier breaks without distractions. You are free to/encouraged to use any AI tools typical to your workflow.

\paragraph{What to build}
\begin{itemize}
    \item A single-page web app with:
    \begin{itemize}
        \item A configurable work timer (e.g., 25 minutes by default)
        \item A break timer that starts automatically after work ends
        \item Start / pause / reset controls
    \end{itemize}
    \item When a break starts, show one mindful activity suggestion (e.g., stretch, breathe, look away from screen)
    \item Persist completed sessions locally in the browser (e.g., localStorage)
    \item Display a simple session history (count or list of completed work sessions)
\end{itemize}
\paragraph{Technical constraints}
\begin{itemize}
    \item Runs fully locally in the browser (open index.html or npm run dev )
    \item Plain JavaScript, or a lightweight framework of choice
    \item Use browser timing APIs ( setInterval , setTimeout )
    \item Local persistence only (e.g., localStorage )
\end{itemize}

\paragraph{Mandatory Requirements}
\begin{itemize}
    \item User can start a work timer and see it count down correctly
    \item When the work timer ends, a break timer begins automatically
    \item A mindful activity suggestion is shown at break start
    \item Refreshing the page does not erase completed session history
    \item The app is usable and testable via localhost or a local HTML file
    \item README to set up and run your code
\end{itemize}

\subsubsection{Task 2: Personalized Recipe Explorer}
Build a local web app that helps users discover recipes based on ingredients they already have and simple preferences. The app queries a public recipe API or LLM and presents a small, usable recipe exploration flow. You are free to/encouraged to use any AI tools typical to your workflow.

\paragraph{What to build}
\begin{itemize}
    \item A web app with:
    \begin{itemize}
        \item An input for available ingredients (free text or comma-separated)
        \item Optional filters (e.g., diet type, max cooking time, calories)
    \end{itemize}
    \item Fetch matching recipes from an LLM or public recipe API
    \item Display a list of results with:
    \begin{itemize}
        \item Recipe title
        \item Key metadata (time, diet tags, calories if available)
    \end{itemize}
    \item Clicking a recipe shows a detail view (ingredients + basic instructions)
    \item Allow users to save favorite recipes locally in the browser
\end{itemize}

\paragraph{Technical constraints}
\begin{itemize}
    \item Runs locally (npm run dev or similar)
    \item Frontend-only UI; API calls may be proxied through a lightweight local server
    \item API key provided via environment variable
    \item Local persistence via localStorage
    \item Responsive layout, but minimal styling is sufficient
\end{itemize}

\subsubsection{Task 3: Digital Vision Board}
Build a local web app where users upload images and arrange them freely on a visual board. The app should support direct manipulation and persistent layout across sessions. You are free to/encouraged to use any AI tools typical to your workflow.

\paragraph{What to build}
\begin{itemize}
    \item A board-style UI where users can:
    \begin{itemize}
        \item Upload images from their computer
        \item Drag, reposition, and resize images on a canvas or grid
    \end{itemize}
    \item Support only image tiles (uploaded images)
    \item Persist board layout and tiles across refreshes
    \item Simple controls to add, move, and delete tiles
\end{itemize}

\paragraph{Technical constraints}
\begin{itemize}
    \item Runs locally (npm run dev or similar)
    \item Images stored locally (e.g., FileReader, Blob, or browser storage)
    \item Live data may be mocked or fetched from a single external API
    \item Persistence via browser storage (e.g., localStorage, IndexedDB)
\end{itemize}

\paragraph{Mandatory Requirements}
\begin{itemize}
    \item User can upload at least one image and see it on the board
    \item Images can be dragged and repositioned
    \item Board state persists after page refresh
    \item At least one non-image tile shows dynamic (changing) data
    \item User can remove a tile from the board
    \item README to set up and run your code
\end{itemize}

\subsubsection{Task 4: Project Planner}
Build a local web app that helps users turn a vague project idea into a concrete plan by breaking it into tasks and visualizing structure over time or hierarchy. You are free to/encouraged to use any AI tools typical to your workflow.

\paragraph{What to build}
\begin{itemize}
    \item A web app where users can:
    \begin{itemize}
        \item Enter a project title and short description
        \item Break the project into tasks and subtasks
    \end{itemize}
    \item Visualize tasks in one structured view:
    \begin{itemize}
        \item Either a task tree (parent / child), or
        \item A timeline-style ordered list
    \end{itemize}
    \item Allow users to:
    \begin{itemize}
        \item Add, edit, and delete tasks
        \item Mark tasks with basic metadata (e.g., status or priority)
    \end{itemize}
    \item Persist the project plan locally in the browser
\end{itemize}

\paragraph{Technical constraints}
\begin{itemize}
    \item Runs locally (npm run dev or similar)
    \item Frontend-only implementation
    \item State and plans persisted via browser storage (e.g., localStorage)
    \item Minimal styling is sufficient; clarity over polish
\end{itemize}

\paragraph{Mandatory Requirements}
\begin{itemize}
    \item User can create a project and add multiple tasks
    \item Tasks can be structured (ordered or hierarchical)
    \item Changes persist after page refresh
    \item User can visually understand task breakdown at a glance
    \item App supports editing and deleting tasks
    \item README to set up and run your code
\end{itemize}

\paragraph{Mandatory Requirements}
\begin{itemize}
    \item User can enter ingredients and retrieve recipe results
    \item Filters affect the returned results correctly
    \item At least one recipe detail page is viewable
    \item Favorite recipes persist across page refreshes
    \item App is usable on a small screen width (basic mobile friendliness)
    \item README to set up and run your code
\end{itemize}

\subsection{Post-Task survey}
\label{sec:app_post_task_survey}

\paragraph{Overview}
After completing the programming task, participants filled out a post-completion survey consisting of (a) a system recall component and (b) self reported measurements on a Likert-scale on various dimensions. We provide questions from the survey here.

\paragraph{System Recall (Representative Questions)}
The system recall section consists of task-specific questions about design and engineering choices which are to be answered in the form of free form text. The full list of questions for each task can be found at \href{https://github.com/vishakhpk/offloading-score}{this linked repository}. Below we provide $2$ representative examples per task.

\textbf{Task 1 (Mindful Break Timer).}
\begin{itemize}
    \item How did you represent the timer state internally? Did you use separate states for work vs. break, or a single timer with mode flags?
    \item How did you persist session history locally? Simple counters, timestamped entries, or structured objects in localStorage?
\end{itemize}

\textbf{Task 3 (Digital Vision Board).}
\begin{itemize}
    \item How did you handle drag-and-drop interactions? Native HTML drag events, pointer/mouse listeners, or a library?
    \item How did you manage image uploads? Base64 encoding, object URLs, or IndexedDB-backed storage?
\end{itemize}

\textbf{Task 4 (Project Planner).}
\begin{itemize}
    \item How did you handle deleting tasks with subtasks? Cascade delete, prevent deletion, or prompt the user?
    \item How did you persist project state locally? A single serialized object or multiple keyed entries in storage?
\end{itemize}

\paragraph{Self-reported Measurements}
The remaining questions are shared across all post-task surveys and capture subjective perceptions of the task and AI usage.

\begin{itemize}
    \item \textbf{Mental demand:} “How mentally demanding was this task?” (7-point scale from not demanding to very demanding, based on cognitive load on the NASA-TLX scale)
    \item \textbf{Trust in AI:} “How much did you trust the output of your AI tools?” (5-point scale)
    \item \textbf{Perceived ownership:} “How much does the final project feel like your output vs. the tool?” (5-point scale)
    \item \textbf{Cognitive contribution:} “What percentage of the thinking did you do yourself vs. the model?” (5-point scale)
    \item \textbf{Reasons for tool use:} What factors result in coding tool use, such as lack of knowledge, speed, habit, curiosity about model outputs, or any other reason (please specify. 
\end{itemize}

\subsection{System Recall Eval}
\label{sec:app_system_recall_eval}
To evaluate system recall, we compared each participant’s answers from the post-task survey about their implementation against the contents of their submitted codebase. For each participant and task, the pipeline evaluates each question individually against the participant’s repository. First, the repository is segmented into text snippets, and for each question-answer pair the system retrieved a small set of candidate snippets using lexical matching over both code content and file-path information. These retrieved snippets, along with the original question and participant answer, were then passed to a \texttt{gpt-5-mini} with a prompt evaluating factual accuracy and completeness relative to the available repository evidence. The judge assigned one of four labels---incorrect, partially correct, mostly correct, or fully correct along with a short evidence-based justification in natural language. We then calculate the average system recall score for each user for the evaluation in \Cref{sec:normative_judgment}.

\paragraph{Prompt for system recall eval.}
\begin{verbatim}
You are grading whether a candidate answer about a code repository 
is supported by the provided repository evidence.
- Use ONLY the provided snippets as evidence.
- Do not speculate beyond the snippets.
- Do not give credit for claims that are plausible but not 
    supported by the evidence.
- Evaluate the answer using both:
    - 1. factual accuracy
    - 2. completeness relative to the question
- Use this 4-level rubric:
    - fully correct: The answer is fully supported by the snippets, 
      materially complete for the question and contains no meaningful
      false, exaggerated, or unsupported claims.
    - mostly correct: The core answer is supported and most 
      important details are correct, but the answer has minor omissions,
      slight overstatement, or a small unsupported detail that does 
      not change the main substance.
    - partially correct: The answer is mixed. Some meaningful parts 
      are supported, but important parts are missing, overgeneralized,
      unsupported, or incorrect.
    - incorrect: The answer is mostly unsupported by the snippets,
      contradicted by the snippets, or wrong on the main point.
- Decision rules:
    - Use fully correct only when essentially all substantive claims 
      are supported by the snippets.
    - If the main answer is right but there are only minor issues, 
      use mostly correct.
    - If support is genuinely mixed on important points, use partially
      correct.
    - If the main takeaway is unsupported or contradicted, use incorrect.
    - A minor unsupported detail should usually prevent fully correct, 
      but may still allow mostly correct.
    - If an unsupported or incorrect detail changes the main meaning 
      of the answer, prefer partially correct or incorrect.
    - When in doubt, prefer the lower label unless the evidence clearly
      supports the higher one.
-In your explanation:
    - cite concrete file paths and line numbers from the snippets
    - explicitly identify which parts of the answer are supported
    - explicitly identify which parts are unsupported or contradicted
    - briefly explain why the chosen label fits better than the 
    neighboring labels
- Return ONLY valid JSON with exactly two keys:
    - "answer"
    - "reason"
- The "answer" value must be exactly one of: '"incorrect", 
    "partially correct", "mostly correct", "fully correct"
\end{verbatim}

\subsection{Code Attribution Evaluation}
\label{sec:app_code_attribution}
We estimate the code attributed to the AI tool using a rule-based pipeline that consumes each participant’s final project codebase with their recorded interaction logs. The primary inputs are the final source files, the raw action trace, and segmented workflow annotations. We first exclude non-informative system artifacts such as lockfiles, dependency trees, and build outputs, from the code base. Then, we extract interaction evidence from the action trace, including recoverable typed text, paste operations, and higher-level annotated episodes such as manual editing, prompt writing, and AI-response review.
Attribution is performed at the file level and then refined at the line level. Human-side evidence comes from direct matches between recovered typed fragments and final code, as well as episodes consistent with manual editing. Tool-side evidence comes from stronger indicators of externally introduced code, such as paste-heavy edits, and annotated episodes from workflow induction suggesting AI-generated changes were applied. The pipeline outputs line- and file-level provenance labels, along with coverage and confidence diagnostics that indicate how much of the codebase could be attributed and how strongly supported those attributions are. The code for this pipeline is made available at \href{https://github.com/vishakhpk/offloading-score}{this linked repository}.
In addition to strict labels, we also calculate a softened attribution view that captures weaker but directional evidence. This version includes code that cannot be assigned with high confidence but is more consistent with being from the tool. Primarily in \Cref{sec:user_study_results}, we report the stricter version of code attribution. In \Cref{sec:app_additional_results}, we confirm that the softened version of code attribution does not change any of the conclusions.

\section{Additional Results}
\label{sec:app_additional_results}

\subsection{Split of \metricname across experiment conditions by task}
\label{sec:app_reliance_by_task}
\begin{figure}
    \centering
    \includegraphics[width=\linewidth]{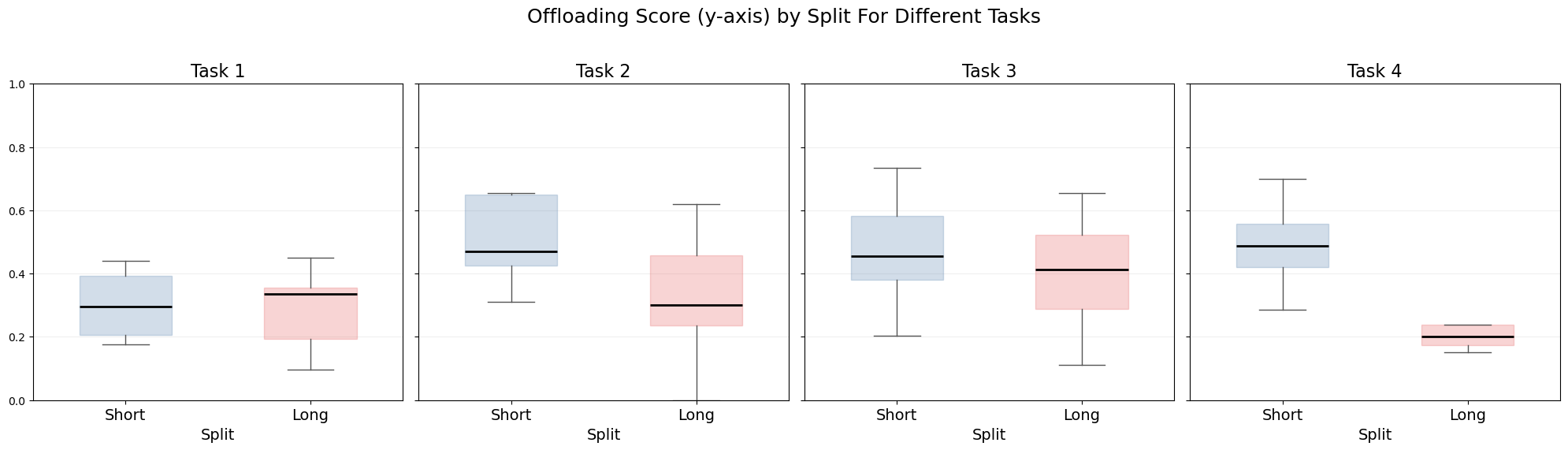}
    \caption{\Metricname values for \shortcond and \longcond conditions for each task. \Metricname assigns lower mean values for each task to varying degrees.}
    \label{fig:offloading_by_task}
\end{figure}

\Cref{fig:offloading_by_task} provides the boxplot of \metricname values for \longcond and \shortcond from \Cref{sec:validation_user_study} when divided into scores by individual task.

\subsection{What fraction of the workflow is associated with \cogprocess and \outputuse interactions?}
\label{sec:app_fraction_labels_workflow}
\begin{figure}
    \centering
    \includegraphics[width=\linewidth]{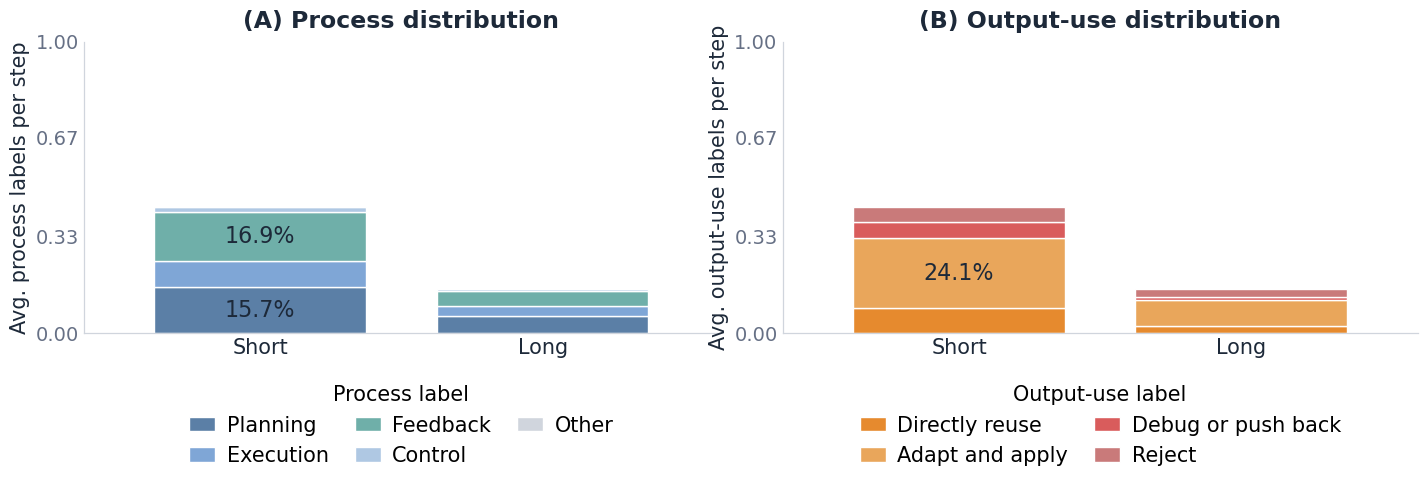}
    \caption{Fraction of workflow steps associated with \cogprocess and \outputuse labels across \shortcond and \longcond conditions. Users in the \shortcond condition directly execute subtasks with the tool 
    and reuse the model outputs, while users in the \longcond condition more frequently reject or adapt outputs, indicating more selective engagement.}
    \label{fig:desc_label_workflow}
\end{figure}

\Cref{fig:desc_label_workflow} is the same interactions as labeled in \Cref{fig:desc_label_dist}, but the denominator to calculate the proportion is the total number of workflow steps. 

\subsection{Do some tools lead to more reliance than others?}
\label{sec:app_reliance_diff_tools}
\begin{figure}
    \centering
    \includegraphics[width=\linewidth]{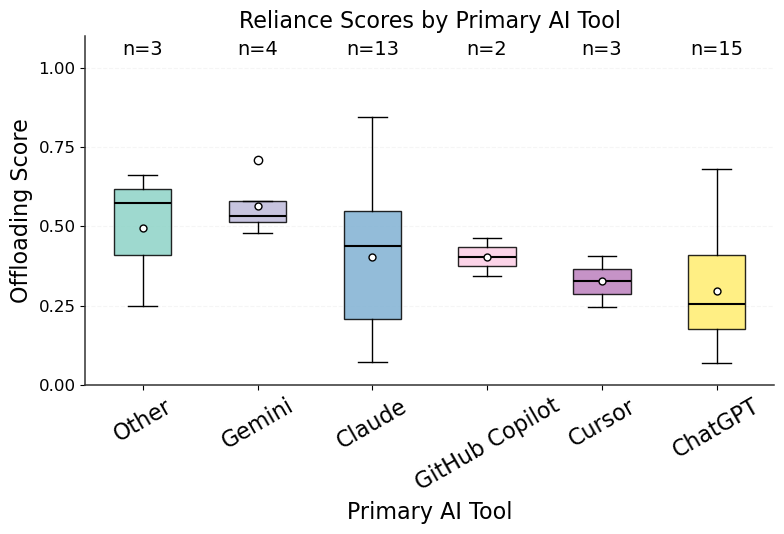}
    \caption{Variation of \metricname values by AI tool.}
    \label{fig:offloading_by_tool}
\end{figure}
We plot the \metricname scores obtained by users using various tools in \Cref{fig:offloading_by_tool}. In general, we don't find a clear pattern of a single tool leading to higher or lower reliance. Most notably, among the most common tools, Claude and ChatGPT, users display a wide range of reliance behavior patterns, indicating that, at this sample size, the human user is the driving factor in the differing reliance patterns, echoing findings from \citet{baumann2026swe}.

\subsection{\Metricname is correlated with first-person perceptions of cognitive offloading and negatively correlated with perceived ownership.} 
\label{sec:app_correlation_judgments}
From the post-completion questionnaire, we calculate the Pearson correlation of \metricname with various dimensions. We find that \metricname is positively correlated with the perceived distribution of cognitive work between the user and tool ($+0.23$) and negatively correlated with perceived ownership of the project ($-0.37$), providing corroborating evidence for validating the metric. Contrary to prior work \citep{merritt2013trust}, we find that trust is only weakly negatively correlated with \metricname ($-0.07$) which may be explained by the observation that users have high trust in coding agents ($35$ out of $40$ participants scored their trust between $4$ and $5$) due to their strong capabilities.

\subsection{Relationship between \cogprocess and \outputuse labels and counterfactual steps.}

\begin{figure}
    \centering
    \includegraphics[width=0.6\linewidth]{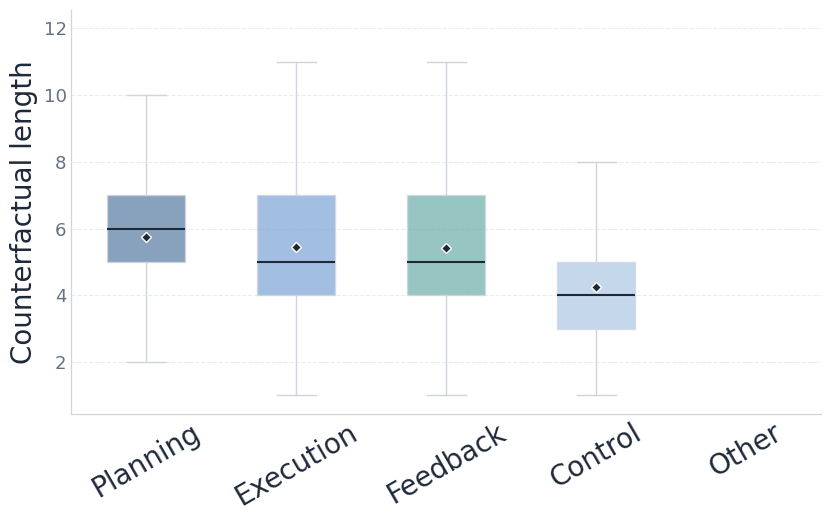}
    \caption{Counterfactual workflow length by \cogprocess label. Planning has a slightly higher median counterfactual length than execution and feedback, while control steps tend to map to shorter counterfactual workflows.}
    \label{fig:counterfactual-length-process}
\end{figure}

\begin{figure}
    \centering
    \includegraphics[width=0.6\linewidth]{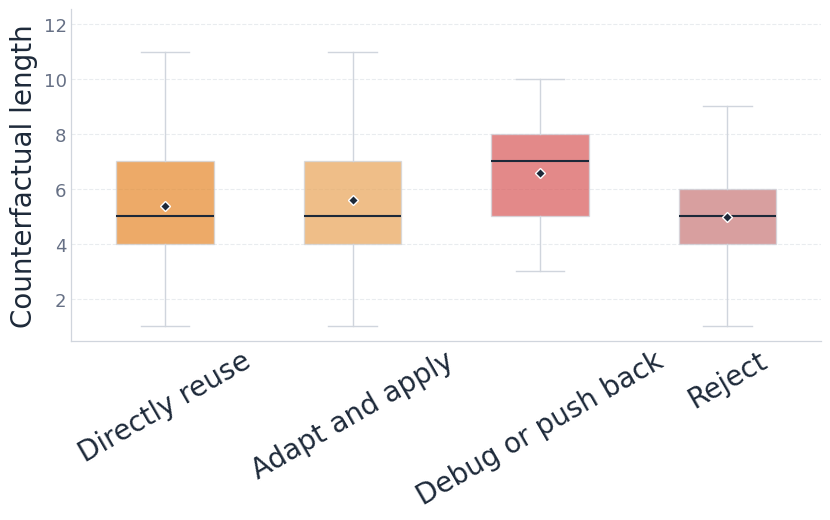}
    \caption{Counterfactual workflow length by \outputuse label. Debug-or-pushback cases have the highest median counterfactual length, suggesting that these interactions correspond to more involved workflow segments.}
    \label{fig:counterfactual-length-output-use}
\end{figure}

\Cref{fig:counterfactual-length-process} shows overlap in counterfactual workflow length across \cogprocess labels, with planning slightly higher than execution and feedback, and control associated with shorter counterfactual workflows. 
\Cref{fig:counterfactual-length-output-use} shows a similar plot across \outputuse labels---debug-or-pushback cases are associated with the longest counterfactual workflows, while directly reuse, adapt-and-apply, and reject have similar, lower medians.

\subsection{Qualitative Interviews for Interpreting User Behavior} 
\label{sec:app_qualitative_interviews}
To better understand these behavioral patterns, we conduct follow-up interviews with two users from each of the three clusters (six total) in \Cref{fig:behavior_patterns_partial}. We first investigate the potentially \emph{overreliant} group, with high reliance and low system recall. $4$ out of $7$ post-task survey responses from this group indicated that they knew how to complete the task independently but chose to use the tool out of habit or to complete the task faster. In follow-up interviews, when reflecting on incorrect answers, users reported that in some cases they were unaware that features had been included in the implementation. In other cases, they had requested a “typical” implementation of a feature and did not verify whether this underspecified intention was followed, reflecting an unresolved grounding gap \citep{shaikh2024grounding}. The second group of users tends to have moderate-to-high reliance and high system recall. In this group, $5$ users reported using their tool for tasks or features for which they did not know how to implement by themselves. In a follow-up interview, one of these users described their process as ``not wanting to limit themselves to one line of thinking'' and hence involving the model in a back-and-forth planning loop. This form of interaction, using their tool to augment their capabilities, is consistent with more \emph{appropriate} reliance in a workflow. Finally, both users interviewed from the lower reliance group reported that a key blocker for them was limited awareness of how the tool could be used effectively. When shown examples of alternative workflows, they indicated that they would have used the tool more if they had known these capabilities, suggesting that lower reliance may reflect a lack of familiarity rather than deliberate avoidance. \footnote{We provide these interview findings as anecdotal qualitative context for the quantitative patterns observed across users, and are not necessarily representative of the full participant pool.}

\subsection{Examples of points of high reliance}
\label{sec:app_example_high_reliance}
We provide two example moments of high reliance below.

\paragraph{Example 1}
The original workflow step:
\begin{verbatim}
- (editing generation) Accept AI-suggested CSS/Controls/Board changes 
for responsive design via multiple “Accept all”/inline Accepts.
\end{verbatim}

And the corresponding counterfactual human-only steps:
\begin{verbatim}
- Open the Implementation Plan / notes that summarize the 
mobile-responsiveness fixes and keep that window visible for reference.
- Use VS Code (or editor) search to find UI/layout sources referenced 
in the notes—open index.css plus component files that contain the 
controls and board layout (e.g., Controls.tsx / Controls.module.css,
Board.tsx / Board.css, App.css).
- Edit index.css to add mobile-targeted rules: add a @media 
(max-width: 600px) block that adjusts root spacing variables, sets the
main container to column flow, reduces paddings/margins and 
font-sizes, and ensures full-width for top-level containers.
- Update controls component styles (Controls.module.css or inline styles
in Controls.tsx) so the control group stacks vertically on small screens
(use flex-direction: column or display: grid with one column) and make 
buttons/inputs expand to 100% width within that breakpoint.
- Adjust board/container styles to preserve aspect and spacing on narrow
viewports (set max-width: 100%, use responsive heights, and ensure 
overflow behavior is handled), and add any utility classes needed to
support the new layout.
- Run the dev server (npm start / yarn start) and open the app in a 
browser; use DevTools Device Toolbar (responsive mode) at common mobile
widths (e.g., 375x667, 414x896) to inspect the layout.
- Interact with the board and controls in the responsive DevTools view, 
noting layout/spacing issues; iterate on the CSS edits (tweak media 
query breakpoints, padding, gaps) until controls and board behave
as expected on small screens.
- Run linting and tests (npm run lint, npm test) and fix any style or 
test failures introduced by the changes (update test snapshots if the 
visual/layout changes require it).
- Create a descriptive commit capturing the changes (e.g., "responsive:
stack controls and add mobile media queries in index.css and component
styles") and push the branch.
\end{verbatim}

\paragraph{Example 2}
The original workflow step:
\begin{verbatim}
- (reading generation) Ask VS Code AI how to add a generate button 
and recipe list using the API key; read its implementation plan.
\end{verbatim}

And the corresponding counterfactual human-only steps:
\begin{verbatim}
- Open the Spoonacular API docs in a browser and locate the recipe search
endpoint (e.g., GET https://api.spoonacular.com/recipes/complexSearch). 
Note required query parameters, response format (where results live), and
how the API key is passed (query param key or header).
- In VS Code open App.jsx and decide where to add UI and state: plan to add
a 'Generate' button, a search input (or reuse existing input value), and a 
recipe list below the input.
- Add React state hooks at the top of the component: recipes (array), loading
(boolean), error (string), and query (string) if not present. 
- Implement a searchRecipes function in App.jsx: set loading true and error 
null; build the request URL using import.meta.env.VITE_SPOONACULAR_KEY and 
the query (and number=5 or similar); call fetch(url), await response.json(),
setRecipes(data.results || data) based on the docs, set loading false; 
catch errors and setError with a readable message and set loading false.
- Wire the UI: add an onChange to the input to update query, add a
<button>Generate</button> with onClick={searchRecipes} (disable it when 
loading), and below add conditional rendering for loading, error, 'no results'
message, and a mapped list of recipes showing title and image (use recipe.title
and recipe.image from the API response).
- Save App.jsx and restart the dev server if necessary (in the terminal 
run npm run dev or stop/start the server) so environment changes (.env.local)
are picked up.', 'Open the app in the browser and test: enter a query, click
Generate, watch the network tab and console for the request/response, 
confirm recipes render correctly; if there are issues (401/403), verify 
the VITE_SPOONACULAR_KEY in .env.local and restart the dev server.
- If the API returns only ids for results, add an extra fetch to the recipe
information endpoint for each id or adjust to call the endpoint that returns
full recipe data; update mapping code to show the additional fields.
\end{verbatim}

\subsection{Example of selective engagement from users in \longcond}
\label{sec:app_example_pushback}
In one \longcond workflow shown below, a participant building a mindful break timer repeatedly used the assistant for design guidance, asking how to redesign timer buttons and session-history cards. Rather than directly accepting the generated changes, the participant reviewed the suggestions, manually edited the prompt and asked for more edits. In a later step, the participant reviewed the AI-proposed UI refactor but did not apply it, instead continuing with manual testing and edits. This illustrates selective engagement: the model is used for planning and guidance, while the user retains control over whether and how outputs enter the implementation.
\begin{verbatim}
- (test manually checking) Run 1-minute work -> 15-second break and
verify suggestions/history updates.
- (reading generation) Review AI notes about card styling 
improvements before applying changes.
- (writing prompt) Send detailed prompt for redesigning the session
history section.
- (writing code) Scroll and update session history CSS while 
referencing AI design notes.
...
- (reading generation) Scroll AI/chat panel to review textual 
description of UI refactor.
- (test manually checking) Run work session through transition 
into break while checking history.
- (writing code) Double-click near bottom of editor to position 
cursor for further HTML edits.
- (reading generation) Clear/collapse AI panel to reset context
while staying on index.html.
- (writing prompt) Submit multi-line prompt requesting updated 
HTML/CSS layout changes.
- (writing code) Scroll/edit CSS, focusing on session-history 
rules while reading AI layout changes.
- (test manually checking) Reload/focus the app, click around 
controls/history, and observe one-minute cycles.
- (writing prompt) Trigger built-in AI chat and request redesign
of break suggestion component.
- (writing code) Position cursor in index.html to review/edit 
CSS for session history/components.
- (test manually checking) Toggle History, clear history, start
work and break sessions, and test Session History modal 
interactions.
\end{verbatim}


\end{document}